\documentclass[english,english,twocolumn,showpacs,preprintnumbers,amsmath,amssymb]{revtex4}
\usepackage[T1]{fontenc}
\usepackage[latin9]{inputenc}
\usepackage{amsmath}
\usepackage{amssymb}
\usepackage{graphicx}
\usepackage{esint}
\usepackage{subfigure}
\usepackage[normalem]{ulem}



\usepackage{babel}
\begin{document}

\title{Linked and knotted chimera filaments in oscillatory systems}

\author{Hon Wai Lau}
\email[]{hwlau@ucalgary.ca}
\affiliation{Institute for Quantum Science and Technology and Department of Physics and Astronomy, University of Calgary, Calgary, Alberta, Canada T2N 1N4}

\author{J\"{o}rn Davidsen}
\email[]{davidsen@phas.ucalgary.ca}
\affiliation{Complexity Science Group, Department of Physics and Astronomy, University of Calgary, Canada T2N 1N4}

\begin{abstract}
While the existence of stable knotted and linked vortex lines has been established in many experimental and theoretical systems, their existence in oscillatory systems and systems with nonlocal coupling has remained elusive. Here, we present strong numerical evidence that stable knots and links such as trefoils and Hopf links do exist in simple, complex, and chaotic oscillatory systems if the coupling between the oscillators is neither too short ranged nor too long ranged.
In this case, effective repulsive forces between vortex lines in knotted and linked structures stabilize curvature-driven shrinkage observed for single vortex rings. In contrast to real fluids and excitable media, the vortex lines correspond to scroll wave chimeras [synchronized scroll waves with spatially extended (tubelike) unsynchronized filaments], a prime example of spontaneous synchrony breaking in systems of identical oscillators.
In the case of complex oscillatory systems, this leads to a novel topological superstructure combining knotted filaments and synchronization defect sheets.
\end{abstract}

%05.45.Xt  Synchronization; coupled oscillators
%89.75.Kd  Patterns (complex systems) 
%82.40.Bj  Oscillations, chaos, and bifurcations
%47.54.-r  Pattern selection; pattern formation (fluid dynamics)

\maketitle

%----------------------------------------------------------------------------------------
\paragraph{Introduction:\label{sec:introduction}}

%The intriguing properties of knots have fascinated scientists for centuries, yet many unresolved questions remain~\cite{prasolov,cromwell}. 
In natural science, knots and linked structures have attracted attention in various branches as they are an essential part of many physical processes. This includes real fluids~\cite{kleckner_creation_2013}, liquid crystals~\cite{machon14,copar15}, Bose-Einstein condensates~\cite{kawaguchi_knots_2008,proment_vortex_2012}, electromagnetic fields and light~\cite{dennis_isolated_2010,kedia_tying_2013}, superconductors~\cite{babaev_non-meissner_2009}, proteins~\cite{wust_sequence_2015} as well as excitable media~\cite{winfree_singular_1983,sutcliffe_stability_2003} and bistable media~\cite{malevanets_links_1996}. Stable knots and their topological invariants are of particular interest for both theory and experiments as they play an important role in characterizing and controlling different systems and their dynamics~\cite{goldstein12}. This is especially true in excitable media, where linked and knotted filaments of phase singularities can be essential to understand the nature of scroll wave propagation processes~\cite{winfree_singular_1983,sutcliffe_stability_2003,kapral,bansagi08}, including nonlinear wave activity associated with ventricular fibrillation and sudden cardiac death~\cite{fenton02,st-yves15}. While the wave propagation processes in excitable and nonlinear oscillatory systems are very similar~\cite{desai}, the existence of such stable knotted and linked filaments in oscillatory systems has remained elusive. For example, to the best of our knowledge no corresponding parameter regime has been identified in the complex Ginsburg-Landau equation (CGLE), which is the normal form of oscillatory media close to the Hopf bifurcation~\cite{aranson_world_2002,rousseau_twisted_2008}. This is deeply unsatisfying as the collective behavior, spontaneous synchronization and wave propagation in oscillatory media and coupled systems of nonlinear oscillators are topics of general interest with applications across disciplines~\cite{pikovsky,desai}, including the quantum regime~\cite{mari_measures_2013,walter_quantum_2014}.

In this paper, we show for the first time that (i) stable knotted and linked filaments do exist in oscillatory systems, (ii) they do exist under non-local coupling in the underlying dynamical equations, and (iii) together with synchronization defect sheets they can form novel topological superstructures. From the Kuramoto model of simple phase oscillators and the CGLE to complex and chaotic oscillatory systems, we find in particular Hopf links and trefoils that persist over hundreds of thousands of scroll wave rotations for a wide range of parameters. 
Due to the non-local coupling, the filaments that make up the long-lived knotted structures are no longer simple phase singularities as is typical for scroll waves, but instead the filaments correspond to spatially extended regions in which the oscillators are unsynchronized. This is despite the fact that all oscillators are identical and uniformly coupled. The coexistence of these unsynchronized local regions with synchronized regions --- exhibiting traveling waves in our specific case --- is the hallmark of a chimera state~\cite{kuramoto02,shima_rotating_2004,abrams_chimera_2004,motter10,panaggio15,maistrenko_chimera_2015}. While single ringlike chimera filaments shrink, knotted and linked filaments generate an effective repulsion that prevents shrinkage and stabilizes the pattern even in the presence of strong noise.
We find that for coupling that is too short ranged (including local coupling) or coupling lag that is too small, the repulsion is too weak such that knotted structures collapse. This is despite the fact that phase twists along the filaments are present, which have been hypothesized to have a stabilizing effect by themselves~\cite{sutcliffe_stability_2003}. If the coupling between oscillators is too long ranged and the coupling lag is too large, straight chimera filaments become unstable in a way reminiscent of negative line tension~\cite{winfree94,biktashev98,alonso03,davidsen08mr,reid09,dierckx12,st-yves15}. This leads to the breakup of the knotted structures as well.

%----------------------------------------------------------------------------------------
\paragraph{Phase oscillators: \label{sec:model}}

As the simplest paradigmatic model of an oscillatory system, we first focus on the Kuramoto model~\cite{kuramoto,acebron_kuramoto_2005,panaggio15}. In this model, $\theta(\mathbf{r},t)\in [-\pi,\pi)$ denotes the state of an oscillator at a spatial point $\mathbf{r}$ and time $t$. The evolution is governed by 
\begin{equation}
\dot{\theta}(\mathbf{r},t)=\omega_0 + K\omega(\mathbf{r},t)\label{eq: Evo}.
\end{equation}
Here, $\omega_0$ is the natural frequency of the oscillators and $K$ is the coupling strength. 
Note that we can set $\omega_0=0$ and rescale time $Kt\to t$ without loss of generality. Thus, $\omega$ is the instantaneous angular frequency obeying 
\begin{equation}
\omega(\mathbf{r},t)=\int_{V}G_{0}(\mathbf{r}-\mathbf{r}')\sin[\theta(\mathbf{r}',t)-\theta(\mathbf{r},t)-\alpha]d\mathbf{r}'
\label{eq:inst_freq}
\end{equation}
where $G_{0}(\mathbf{r})$ is a coupling kernel, $\alpha$ is the
coupling lag or phase shift, and the integration is taken over the whole volume $V$.
The kernel used is a top-hat kernel with coupling radius $R$ 
\begin{equation}
G_{0}(\mathbf{r})\sim\begin{cases}
1, & r\le R\\
0, & r>R
\end{cases}
\label{eq:kernel}
\end{equation}
which is normalized as $\int_{V}G_{0}(\mathbf{r})d\mathbf{r}=1$.
In simulations, the spatial locations are discretized into $\mathbf{r}=(x_{i},y_{i},z_{i})$
with $1\le x_{i},y_{i},z_{i}\le L$ taking integer values in the system
of linear length $L$ such that $V=L^3$. Hence, the control parameters of the system are $R$ and $\alpha$, with finite size effects determined by $L$. Extensive simulations have been done using the Runge-Kutta scheme
\footnote{
We have tested different time steps using both explicit Runge-Kutta and Euler's method. The stable topological structures are preserved but the trajectories deviate after a long time. 
This is expected since a large time step introduces an effective noise. Similar observations have been mentioned in Ref.~\cite{sutcliffe_stability_2003} for excitable media.
} with both random initial conditions (IC) and specific functions, see the Supplemental Material (SM) for details~\cite{supplementary}.
We use periodic boundary conditions (BC) here, yet our findings are quite independent of the BC.

The Kuramoto model with nonlocal coupling is known to exhibit chimera states, in which both synchronized and unsynchronized regions of oscillators can coexist in the same system even though all oscillators are identical and uniformly coupled.
Most studies have been done on the one dimensional ring and complex networks~\cite{panaggio15}.
In higher dimensions, two qualitatively different chimera regimes have been identified for the Kuramoto model given by Eqs.~\eqref{eq: Evo},~\eqref{eq:inst_freq},~\eqref{eq:kernel}.
For near global coupling with $R\sim L$ and large $\alpha \lesssim \pi/2$, various coherent and incoherent strip, spot, plane, cylinder, sphere and cross patterns have been observed in two and three dimensions (2D and 3D)~\cite{omelchenko_stationary_2012,maistrenko_chimera_2015}. 
The other regime involves shorter range nonlocal coupling $L\gg R\gg 1$ with smaller $\alpha$.
In 2D geometries, synchronized spiral waves with unsynchronized chimera cores can appear. They behave like a normal spiral, yet the dynamics in the core is unsynchronized~\cite{kim_pattern_2004,martens10,omelchenko_stationary_2012,panaggio_chimera_2015}. Similarly, in 3D, regular scroll waves with chimera filaments (or chimera tubes) at their center --- instead of the linelike filaments of phase singularity --- have been observed~\cite{maistrenko_chimera_2015}.

%----------------------------------------------------------------------------------------
\paragraph{Existence of knots: \label{sec:results}}

\begin{figure}
\begin{centering}
\includegraphics[width=.98\columnwidth]{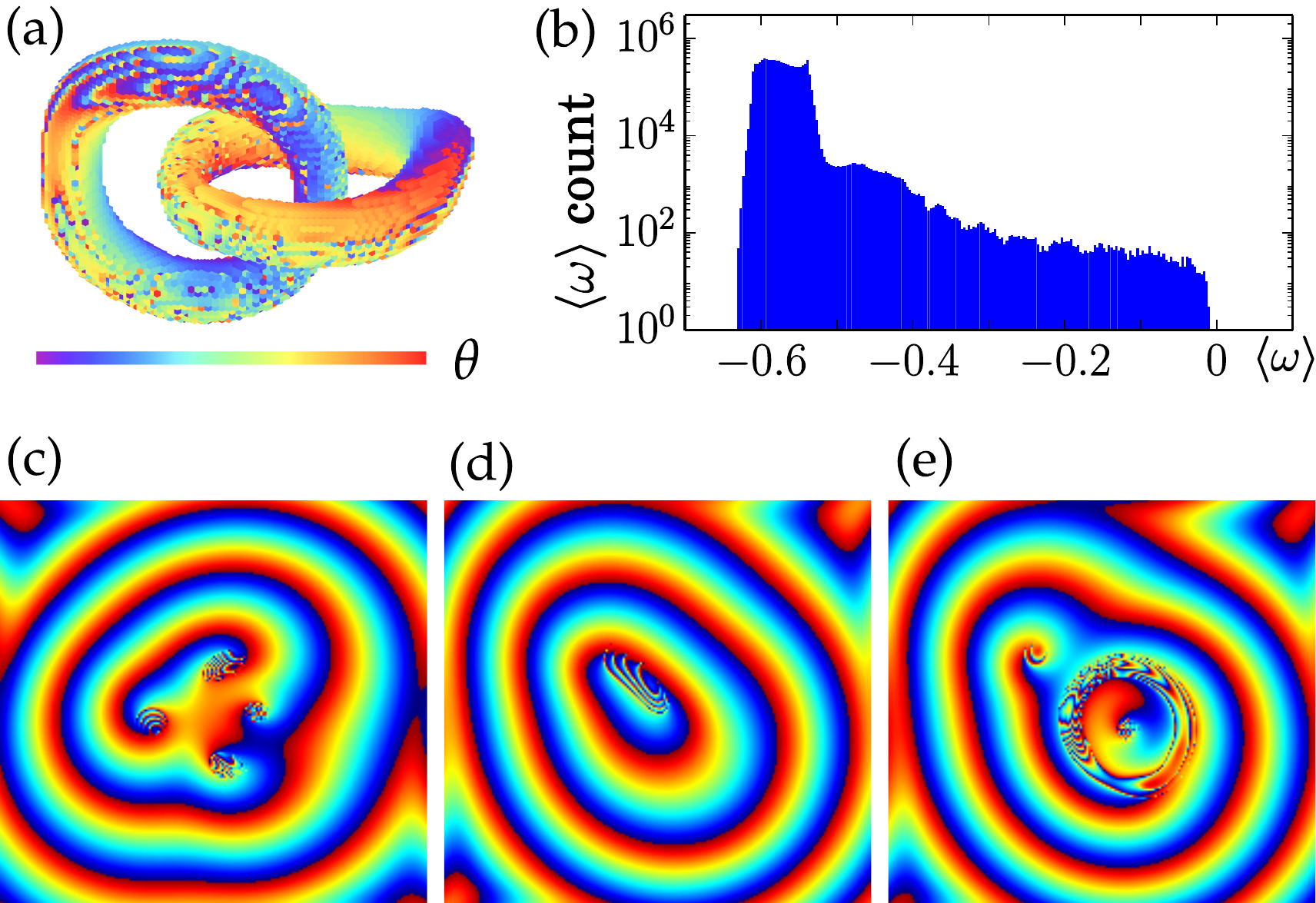}
\par\end{centering}

\caption{\label{cap: chimera-properties} (color online) Chimera nature of knots ($\alpha=0.8$, $R=8$, $L=200$). (a) Snapshot of a Hopf-link shows the unsynchronized phase $\theta(\mathbf{r},t)$ on the isosurface of spatially smoothed angular frequency $\tilde{\omega}(\mathbf{r},t)$. 
(b) Mean angular frequency distribution of $\langle	\omega(\mathbf{r}) \rangle$, averaged over approximately 10 periods. The tail to the right corresponds to unsynchronized oscillators. (c) Plot of 2D cross-section of $\theta(\mathbf{r})$ showing the chimera and spiral wave properties. A vertical cut through both rings in panel (a), showing four chimera cores. (d) A cut through the far edge of a ring. (e) Slicing of a ring, showing a ring chimera and two chimera cores of the other ring.
}
\end{figure}

\begin{figure}
\begin{centering}
\includegraphics[width=0.98\columnwidth]{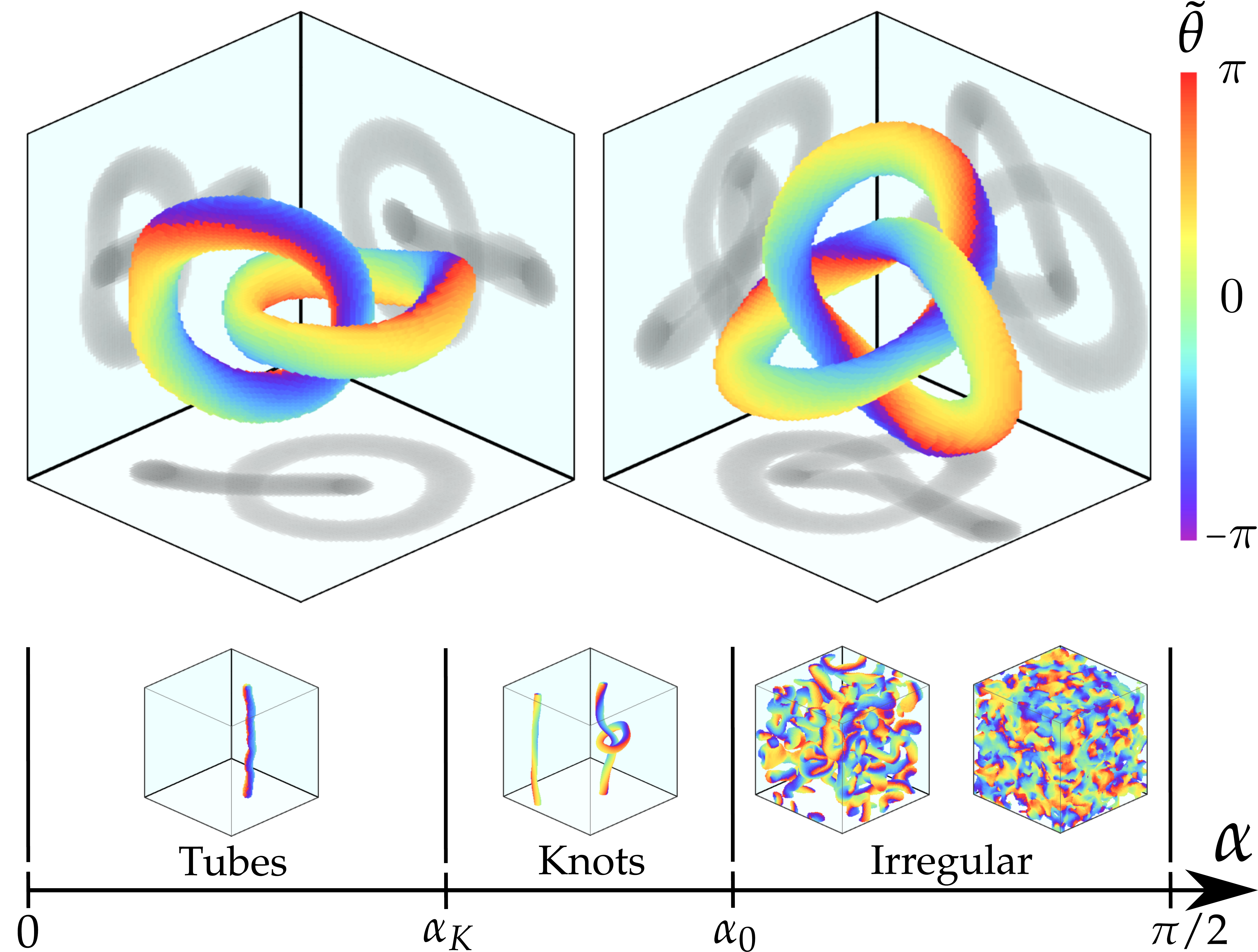}
\par\end{centering}

\caption{\label{cap: phase-diagram} (color online) Phase diagram of the
Kuramoto model in 3D as a function of $\alpha$ with nonlocal coupling
$L \gg R \gg 1$. Various knots exist between $\alpha_K$ and
$\alpha_0$. Top panel shows stable Hopf-link (left), and trefoil (right) as examples.
The plots are similar to those in Fig.~\ref{cap: chimera-properties}a, but smoothed phases $\tilde{\theta}(\mathbf{r})$ are used instead. 
Shadows on the walls correspond to perpendicular projections of the structures.
}
\end{figure}

For $L\gg R\gg 1$ and large effective system size $L/R$,
we observe different stable linked and knotted scroll waves in the Kuramoto model as shown in Figs.~\ref{cap: chimera-properties} and ~\ref{cap: phase-diagram}. To clearly visualize the chimera tubes and the knotted and linked structures (referred to as knots in the following), one has to take into account that both phase $\theta(\mathbf{r},t)$ and angular frequency $\omega(\mathbf{r},t)$ fluctuate a lot in space as shown in Fig.~\ref{cap: chimera-properties}b-~\ref{cap: chimera-properties}e. Thus, it is helpful to define a local mean angular frequency $ \tilde{\omega}(\mathbf{r},t)=\int_{V}G_{0} (\mathbf{r}-\mathbf{r}')\omega(\mathbf{r}',t)d\mathbf{r} $. Fig.~\ref{cap: chimera-properties}a shows a snapshot of the chimera tubes by plotting the phases of the unsynchronized oscillators for $\tilde{\omega}(\mathbf{r},t) \ge \text{const}$. The presence of scroll waves with chimera filaments can also be seen directly in the phase field. Selected 2D cross-sections of the phase field (Fig. \ref{cap: chimera-properties}c) show patterns similar to chimera spirals in 2D~\cite{kim_pattern_2004}, while other cross-sections show features that are specific to 3D such as the chimera ring shown in Fig.~\ref{cap: chimera-properties}e.
Note that the Hopf link and other knots observed generate spherical wave in the far-field. Moreover, they are not stationary but keep rotating, drifting, and changing their shape over time as shown in the videos and the SM~\cite{supplementary}.

As chimera filaments are associated with scroll waves, phase twists can be present along the filament \cite{winfree_stable_1990}. This is visible in Fig.~\ref{cap: chimera-properties}a, but can be better visualized by considering the local mean phase field $\tilde{\theta}(\mathbf{r},t)$, defined by
\begin{equation}
\tilde{\rho}(\mathbf{r},t)e^{i\tilde{\theta}(\mathbf{r},t)}=\int_{V}G_{0}(\mathbf{r}-\mathbf{r}')e^{i\theta(\mathbf{r}',t)}d\mathbf{r}'.
\label{eq:mean_phase}
\end{equation}
This is illustrated in Fig.~\ref{cap: phase-diagram}, for example.

%----------------------------------------------------------------------------------------
\paragraph{Phase diagram: \label{sec:phase}}

\begin{figure}
\begin{centering}
\includegraphics[width=0.98\columnwidth]{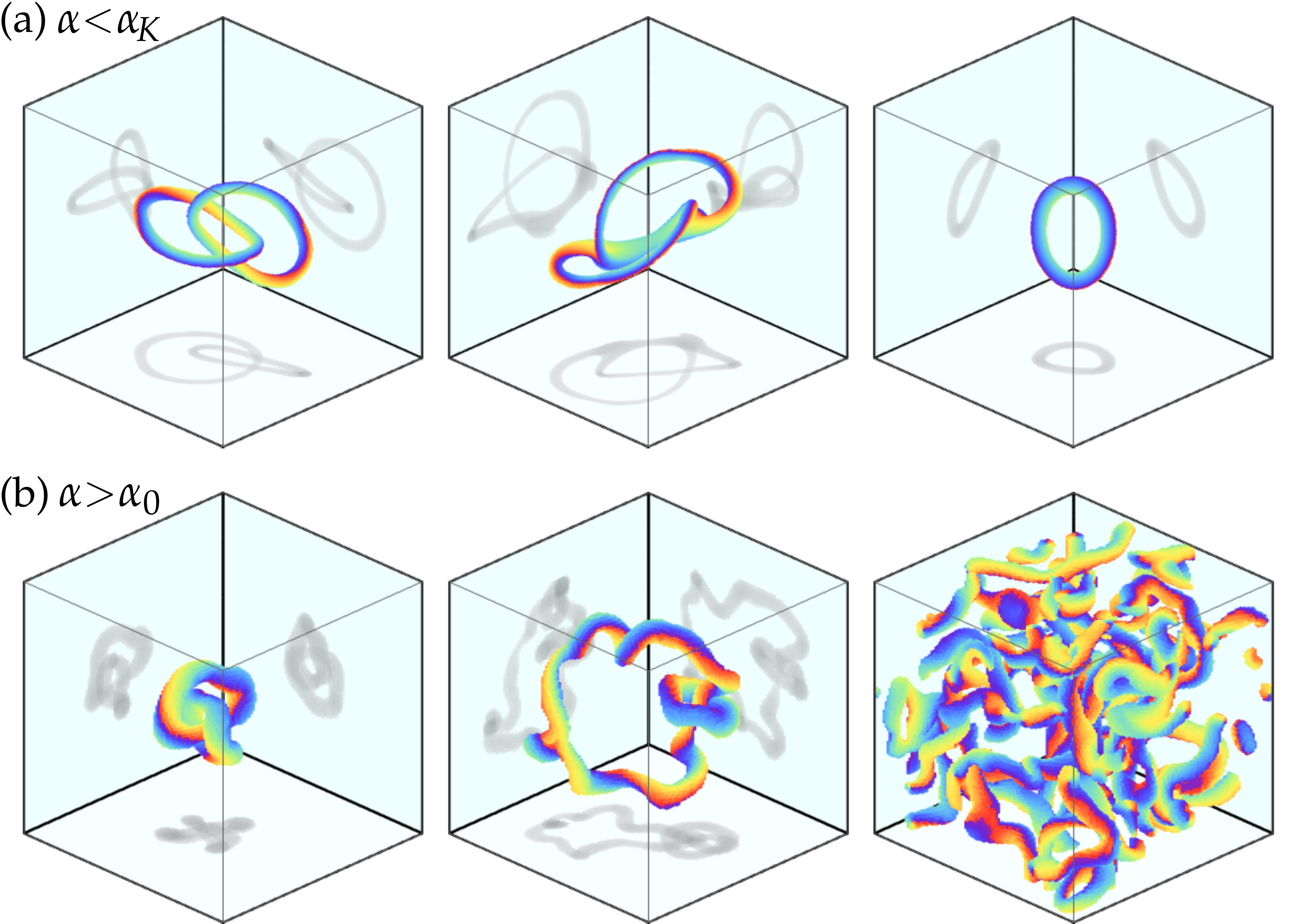}
\par\end{centering}

\caption{\label{cap: near-phase-transition} (color online) Time evolution of Hopf links outside their stability regimes near $\alpha_K$ and $\alpha_0$, respectively. (a) $\alpha< \alpha_K$. After $\alpha$ is decreased slightly from above $\alpha_K$ to below, the two rings of the Hopf link merge resulting in a single ring which eventually vanishes. (b) $\alpha>\alpha_0$. After $\alpha$ is tuned slightly from below $\alpha_0$ to above, one of the rings grows until it collides and reconnects, resulting in a turbulent-like pattern.
}
\end{figure}

For the Kuramoto model given by Eqs.~\eqref{eq: Evo},~\eqref{eq:inst_freq}, and~\eqref{eq:kernel}, our numerical simulations allow us to obtain a phase diagram as a function of $\alpha$. This is plotted in Fig.~\ref{cap: phase-diagram} together with some of the asymptotic states. At small $\alpha$, only relatively simple scroll wave structures with straight chimera tubes are stable. For $\alpha_K < \alpha < \alpha_0$, also knots such as 1 twist Hopf links and 3 twist trefoils (as shown in Fig.~\ref{cap: phase-diagram}) are stable over hundreds of thousands of scroll wave rotations period $T$
\footnote{The stability of Hopf links and trefoils in the Kuramoto model has been tested for extended periods of time of at least $t=1.2\times 10^6$ (or period $T>10^5$ where $T\approx 11$ at $\alpha=0.8$) for $L=100$ with $R=4$, and $t=1.2\times 10^5$ for $L=200$ with both
$R=4$ and $R=8$.}. 
More stable structures including helices, ring-tubes and linked triple rings are shown in the SM~\cite{supplementary}. For $\alpha > \alpha_0$, knots as well as simple straight tubes become unstable. The evolution in the former case is shown in Fig.~\ref{cap: near-phase-transition}b. In the latter case, the dynamics of the chimera filament indicates that a finite wavelength instability of the filament itself occurs such that the filament grows rapidly (see the SM~\cite{supplementary}). In both cases, the rapid growth of filaments is accompanied by fragmentation through collisions leading eventually to an irregular or turbulent-like behavior as shown in Fig.~\ref{cap: phase-diagram}. Furthermore, in the same parameter regime near $\alpha_0$ in 2D, chimera spirals are stable and no irregular pattern is present~\cite{kim_pattern_2004}. All this suggests that the underlying instability is truly 3D in nature as the negative line tension instability and similar filament instabilities that have been observed in excitable and oscillatory media~\cite{st-yves15}. 
Fig.~\ref{cap: phase-diagram} also shows that at even higher $\alpha \sim \pi/2$, no filament structures can be recognized.

The nature of the instability of knots at $\alpha_K$ and $\alpha_0$ are significantly different. Below $\alpha_K$, any knot transforms through one or multiple reconnections into a single untwisted ring which shrinks and disappears, leading to homogeneous oscillations. As we have tested, all single chimera rings with a radius of up to 80 shrink and eventually vanish for $0 \leq \alpha<\alpha_0$ with no-flux BC (see SM~\cite{supplementary}). This together with the stable knots for $\alpha_K < \alpha < \alpha_0$ indicates that there is an effective repulsion between filaments in knots that is sufficient to prevent curvature-driven shrinkage and stabilize these structures above $\alpha_K$. Below $\alpha_K$, the repulsion is too weak to prevent reconnections. This mechanism is similar to what has been observed for knots in bistable media~\cite{malevanets_links_1996} and plays an important role in other situations as well~\cite{jimenez12}.
Simulation results show that different knots have different stability regimes, especially Hopf links are stable over a broader range of $\alpha$ than trefoils. Therefore, we denote $\alpha_K$ in the following as the point at which Hopf links disappear.

%----------------------------------------------------------------------------------------
\paragraph{Dependence on $R$, $L$, and geometry: \label{sec:r}}

Numerical simulations for $4 \leq R \leq 12$ and $64 \leq L \leq 300$ show that the phase diagram presented in Fig.~\ref{cap: phase-diagram} is independent of the specific choice of $R$ and $L$ as long as $L\gg R\gg 1$. 
Specifically, $\alpha_K \approx 0.61$ and $\alpha_0 \approx 0.90$ with uncertainty $\pm 0.02$.
The condition $L\gg R$ ensures that finite size effects do not play a significant role as the size of stable knots and the wavelength scale with $R$~\cite{martens10,kim_pattern_2004}. For example, we find that stable Hopf links cease to exist for $L/R \lesssim 16$. Also, if the effective system size is too small, more complex knots tend to decay into simpler ones (see the SM~\cite{supplementary}). 
The condition $R \gg 1$ is also crucial. We find that for shorter range couplings $R<3$ the lifetime of knots is finite
\footnote{For $R=2$, the lifetime can vary significantly with the used time stepping of the integrator and becomes longer for shorter $\Delta t$. For example, the lifetime is about $t = \mathcal{O}(5000)$ and, thus, less than 1000 scroll wave rotations using 4th order Runge-Kutta with $\Delta t=0.02$.}. 
Specifically, no stable knots have been observed for local coupling, $R=1$, independent of the IC used to generate Hopf links.
This is a consequence of temporal fluctuations in the s
hape of the individual rings within a Hopf link becoming comparable to the minimum separation between the rings 
%, similar to the illustration in Fig.~\ref{cap: cgle-rossler}a, 
such that the rings merge and disappear (see the SM~\cite{supplementary}) --- the same behavior as for the instability at $\alpha_K$. We observe qualitatively the same for trefoils.

%----------------------------------------------------------------------------------------
\paragraph{Robustness with respect to noise: \label{sec:noise}}

\begin{figure}
\begin{centering}
\includegraphics[width=.48\columnwidth]{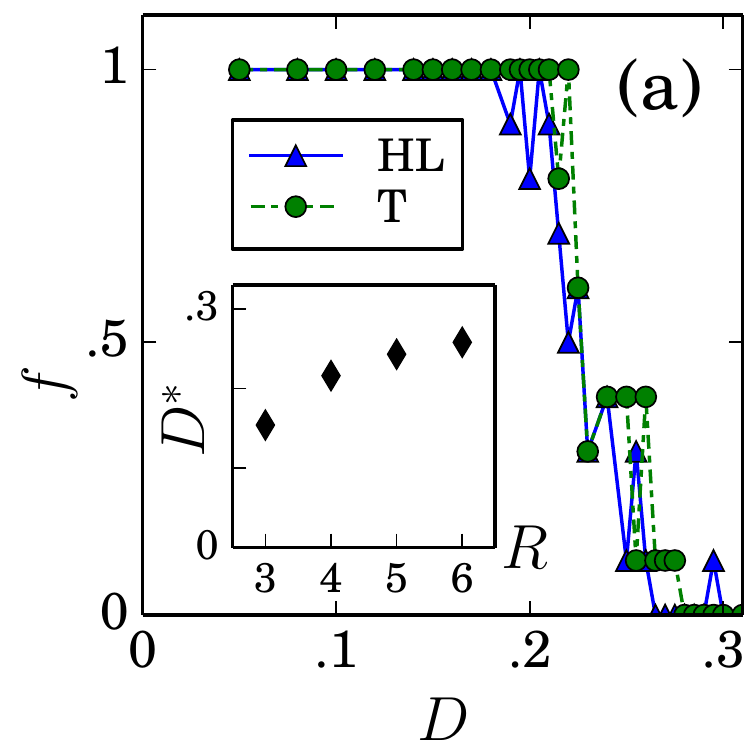}~~~~\includegraphics[width=.48\columnwidth]{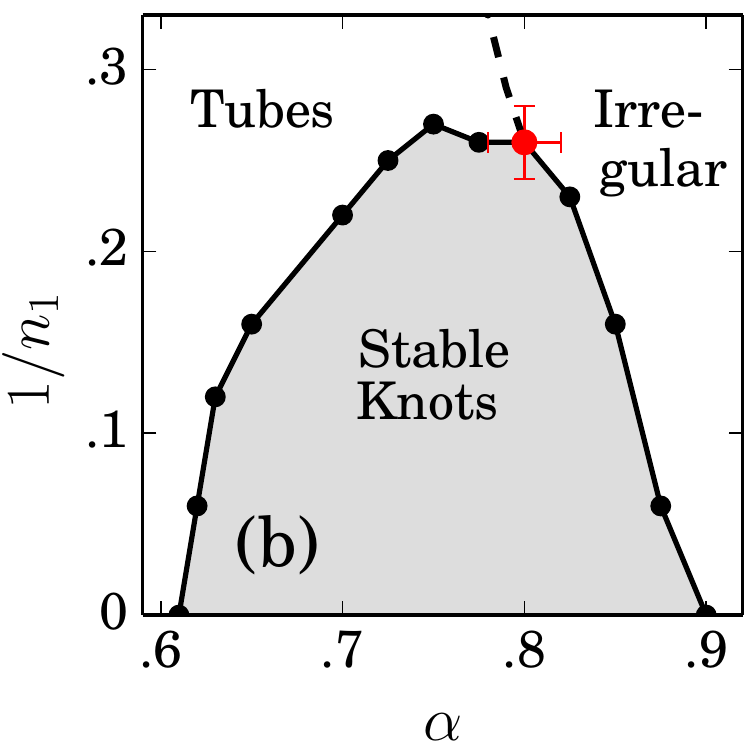}
\par\end{centering}

\caption{\label{cap: noise-tolerance} (color online) (a) The tolerable phase noise level for Hopf links (HL) and trefoils (T). $f$ quantifies the fraction of structures persisting after $t=5000$ ($\approx 450$ scroll wave rotations) under noise intensity $D$. Each point corresponds to an ensemble of 10 realizations. Here, $\alpha=0.8$, $R=4$, and $L=100$.
Inset: $D^*$ denotes the point of $f=0.5$ for a Hopf link as a function of $R$ using $\alpha=0.8$, $L/R=16$.
(b) Phase diagram showing the stable regime of Hopf links for kernel $G_1$ with $L\gg R\gg 1$. The red dot marks the triple point between all three phases where $\alpha_K = \alpha_0$. The error bars account for different $R \geq 4$ and $L$ up to $L=300$. Note that $1/n_1 = 0$ corresponds to the top-hat kernel $G_0$ used in Fig. \ref{cap: phase-diagram}.
}
\end{figure}

To further quantify the stability of different topological states, we examine them in noisy environments. This is modeled by an additional Gaussian phase noise $\xi(\mathbf{r},t)$ in the Kuramoto model 
\begin{equation}
\dot{\theta}(\mathbf{r},t)=\omega(\mathbf{r},t)+D\xi(\mathbf{r},t)
\end{equation}
where $\langle\xi(\mathbf{r},t)\rangle=0$ and $\langle\xi(\mathbf{r},t)\xi(\mathbf{r}',t')\rangle=\delta(\mathbf{r}-\mathbf{r}')\delta(t-t')$.
As shown in Fig. \ref{cap: noise-tolerance}, Hopf links and trefoils can survive under noise magnitude as high as $D^*=0.22$.
% or $|D^*/\langle\omega\rangle|=0.38$.
This high robustness under noise signifies the topological protection of knots. Longer range coupling as quantified by $R$ also increases the tolerance of local phase noise as shown in Fig. \ref{cap: noise-tolerance}b.

%----------------------------------------------------------------------------------------
\paragraph{Dependence on spatial kernel: \label{sec:kernel}}

In contrast to the top-hat kernel $G_0$, we did not observe stable knots for Gaussian kernels often considered in the context of chimera states. 
This together with the existence of a minimal $R$ discussed above indicates that the range of the spatial kernel is crucial. To substantiate this further, let us consider
the kernel $G_{1}(r) \sim e^{-(r/R){}^{n_{1}}}$ such that $G_{1}\to G_{0}$ when $n_{1}\to\infty$ if $R<L$.
Note that $n_{1}=2$ is a Gaussian, $n_{1}=1$ is an exponential, and $n_1=0$ gives global coupling. Simulations show that if $G_{1}$ becomes more long-ranged as $n_1$ decreases, the stable regime $\alpha_K < \alpha < \alpha_0$ of knots shrinks as shown in Fig.~\ref{cap: noise-tolerance}c. This is the only effect on the knots as the nature of the associated instabilities along the boundaries appears unchanged and follows the pattern shown in Fig.~\ref{cap: near-phase-transition}.
The phenomenon is independent of the exact functional form of the kernel (see the SM~\cite{supplementary}).

%----------------------------------------------------------------------------------------

\paragraph{Beyond phase oscillators: \label{sec:cgle}}
  
\begin{figure}
\begin{centering}
\includegraphics[width=.48\columnwidth]{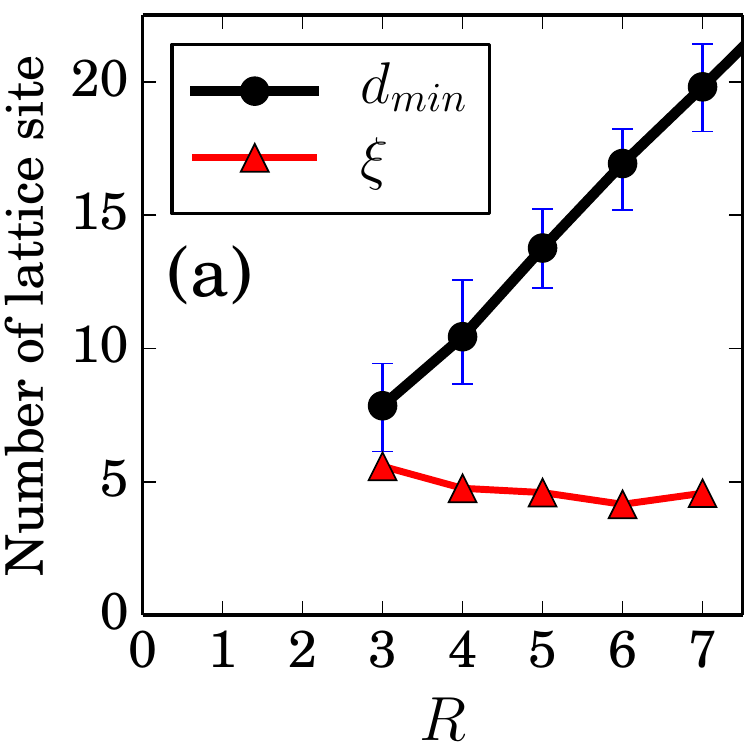}~~\includegraphics[width=.48\columnwidth]{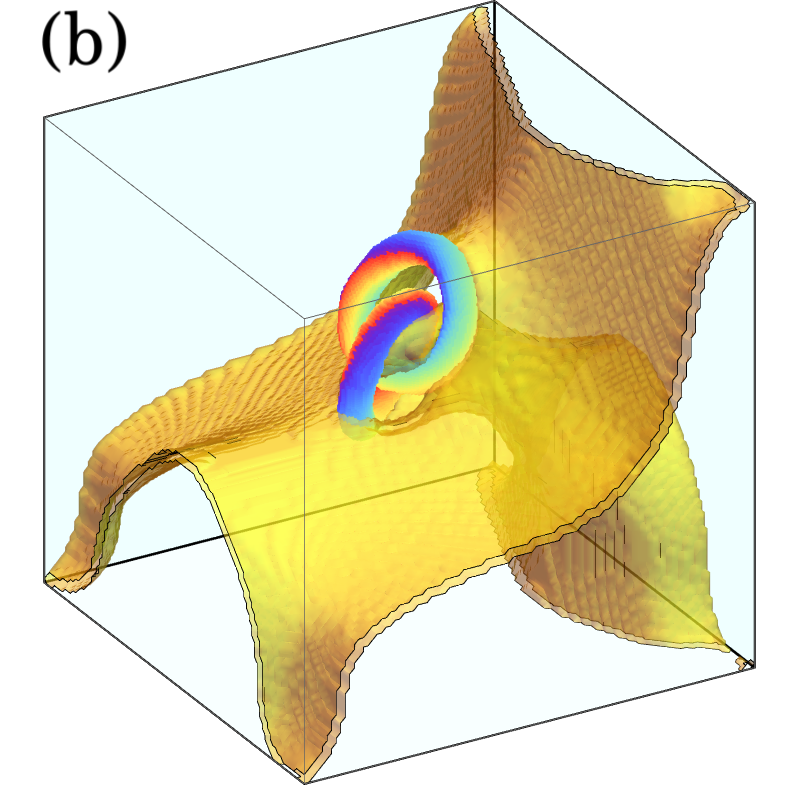}
\par\end{centering}

\caption{\label{cap: cgle-rossler}(color online) (a) CGLE [$(a,b)=(1,0)$, $K=0.1$]: Comparison of minimum separation between rings in a Hopf link, $d_{min}$, and a measure of spontaneous fluctuations in the ring shape, $\xi$ (see the SM~\cite{supplementary} for details). Bars indicate the 99\% range.
%range of values of $d_{min}$ over $t=25000$. 
(b) R\"ossler model: Topological superstructure of a chimera Hopf link with an attached sychronization defect sheet in the period doubled regime (see the SM~\cite{supplementary} for more viewpoints and dynamics). 
}
\end{figure}

The nonlocal CGLE is given by~\cite{kuramoto_rotating_2003}:
\begin{equation}
\dot{A}(\mathbf{r},t)=A-(1+ib)|A|^{2}A+(1+ia) p_A,
\end{equation}
where the control parameters are $(a,b)$. The nonlocal coupling $p_A$ is given by
\begin{equation}
p_A(\mathbf{r},t) = K \int G_0(\mathbf{r}-\mathbf{r'})[A(\mathbf{r}',t)-A(\mathbf{r},t)]d\mathbf{r}',
\label{eq:cgle_kernel}
\end{equation}
and the coupling strength is $K$. Since the CGLE can be well approximated by the Kuramoto model in the weak-coupling limit independent of the specific coupling~\cite{gu,shima_rotating_2004}, similar results are expected in certain parameter regimes. Indeed, stable chimera knots with \emph{non-constant} amplitudes $|A|$ exist in the vicinity of the parameters $(a,b)=(1,0)$~\cite{kuramoto_rotating_2003} for $K=0.1$ (see the SM~\cite{supplementary}) and larger values of $K$. The lifetimes of knots are longer than $6 \times 10^5$, provided that $R \gg 1$. All results discussed above for the Kuramoto model also hold qualitatively for the CGLE. This includes in particular the break-up of knots for small $R$. Fig.~\ref{cap: cgle-rossler}a provides a clear rationale why this happens: The separation between the rings in a Hopf link shrinks with decreasing $R$ such that it eventually becomes comparable to the amplitude associated with the temporal fluctuations in the shape of individual rings. This offers an explanation of why no stable knots have been observed in the CGLE with local coupling.
Our findings for the CGLE imply that \emph{all} oscillatory systems with appropriate nonlocal coupling should exhibit stable knots in some parameter regime near their Hopf bifurcation.

%

%----------------------------------------------------------------------------------------

\paragraph{Complex oscillatory systems: \label{sec:rossler}}
We also observe stable chimera knots if the uncoupled oscillators are far from the Hopf bifurcation and undergo complex or even chaotic oscillations, requiring at least a three-dimensional local phase space. A specific example is the R{\"o}ssler model with nonlocal coupling~\cite{gu_spiral_2013}, which exhibits a phenomenology with many features in common with those observed in complex oscillatory systems including chemical experiments~\cite{davidsen04h}. It is given by 
\begin{eqnarray}
\dot{X}(\mathbf{r},t) & = & -Y-Z+p_X,\nonumber\\
\dot{Y}(\mathbf{r},t) & = & X+aY+p_Y,\\
\dot{Z}(\mathbf{r},t) & = & b+Z(X-c),\nonumber
\end{eqnarray}
where the control parameters are $(a,b,c)$ and the nonlocal coupling $p_X(\mathbf{r},t)$ and $p_Y(\mathbf{r},t)$ are defined analogously to Eq.~\eqref{eq:cgle_kernel}. 
For $a=b=0.2$, the effective $\alpha$ decreases as $c$ increases~\cite{gu}. For $R \gg 1$, we observe stable chimera knots in the period-doubled regime ($c=3.6$) and in the chaotic regime ($c=4.8$) with weak coupling $K=0.05$ (see the SM~\cite{supplementary}).
All findings described above for the other models hold qualitatively here as well. Stable knots only exist if the coupling between the oscillators is neither too short-ranged nor too long-ranged.
For example, we did not observe stable Hopf links or trefoils for $R=1$ or when the kernels were Gaussian in the parameter regimes given above.
Moreover, in the period-doubled regime, synchronization defect sheets (SDS) --- the analog of synchronization defect lines in 2D systems~\cite{goryachev00,gu_spiral_2013} --- can be observed for the first time and, more importantly, connect the different filaments (see Fig. \ref{cap: cgle-rossler}b, the SM and movie~\cite{supplementary}). This leads to another layer of topological structure associated with the knots, making this a unique phenomenon and adding potentially to their general robustness if multiple knots are present~\cite{davidsen04h}.
%

%----------------------------------------------------------------------------------------
\paragraph{Discussion and conclusions: \label{sec:discussion}}

Our findings show that knots exist and are stable over a significant range of parameters in various oscillatory systems with nonlocal coupling as long as the characteristic coupling length of the kernel is sufficiently large and the tail of the kernel decays sufficiently fast.
The variety of knots is also much higher compared to what has been reported for excitable media~\cite{winfree_stable_1990,sutcliffe_stability_2003}. For example, we have also observed other relevant unknotted structures such as stable double helices (see the SM~\cite{supplementary}) --- a structure that has remained elusive in the CGLE with local coupling~\cite{aranson_dynamics_1998}. This suggests that the models considered here can serve as paradigmatic models to study various knotted and unknotted structures associated with scroll waves in general, including the novel topological superstructures of knots with SDSs. More specifically, it allows one to explore the topological constraints imposed by the phase field on the observable phase twists associated with a given knot --- a field largely untouched~\cite{winfree_stable_1990} --- as well as the effect of synchronization defect sheets on knots for the first time.

A remaining open question is to which extent the existence of stable knots in oscillatory systems depends on the presence of a chimera state. While our findings suggest that a chimera state is a necessary condition, there is no fundamental reason to substantiate this. However, our simulations indicate that the mobility of the scroll wave filaments plays an important role. If the filaments move or meander sufficiently fast (e.g. $R=1$ or for a Gaussian kernel with large $\alpha$), no chimera state can be numerically observed and stable knots are absent. This is similar to what has been reported for chimera spirals in 2D~\cite{martens10} and knots in excitable media~\cite{sutcliffe_stability_2003}.
One possible way forward is the recently proposed ansatz by Ott and Antonsen~\cite{ott_low_2008,ott_long_2009}, which has been successfully applied to study the existence and stability of chimera spirals~\cite{laing_dynamics_2009}.

In addition to the robustness of knots under dynamical noise, we also find that Hopf links and trefoils can emerge in a self-organized way from random IC with fair probability (see the SM~\cite{supplementary}). Both features indicate knots should be observable in real-world oscillatory systems that follow a dynamics similar to the models studied here, with most likely candidates to be chemical systems~\cite{shima_rotating_2004,gu_spiral_2013,jimenez12,kupitz13,nkomo13}.
Yet, the observation of chimera filaments in natural systems remains a challenge for the future.

\textit{Acknowledgments}: We thank G. St-Yves, Y. Maistrenko, and V. Maistrenko for helpful discussions.
This research was enabled in part by support provided by WestGrid, Calcul Qu{\'e}bec, and Compute Canada.
H.L. was financially supported by AITF.
J.D. was financially supported by NSERC.

\bibliographystyle{apsrev}
%\bibliography{j2,ref}

%%%%%%%%%% Merge with supplemental materials %%%%%%%%%%
\pagebreak
\clearpage
\widetext
\begin{center}
\textbf{\large Supplemental Materials: Linked and knotted chimera filaments in oscillatory systems}
\end{center}
%%%%%%%%%% Merge with supplemental materials %%%%%%%%%%
%%%%%%%%%% Prefix a "S" to all equations, figures, tables and reset the counter %%%%%%%%%%
\setcounter{equation}{0}
\setcounter{figure}{0}
\setcounter{table}{0}
\setcounter{page}{1}
\makeatletter
\renewcommand{\theequation}{S\arabic{equation}}
\renewcommand{\thefigure}{S\arabic{figure}}
\renewcommand{\bibnumfmt}[1]{[S#1]}
\renewcommand{\citenumfont}[1]{S#1}
%%%%%%%%%% Prefix a "S" to all equations, figures, tables and reset the counter %%%%%%%%%%

\newcommand{\lyxdot}{.}

\section{Topological structures}

\begin{figure*}[p]
\begin{centering}
\subfigure[$\tau>1.2\times10^{6}$, $\alpha=0.8$, $R=4$]{\begin{centering}
\includegraphics[width=0.32\columnwidth]{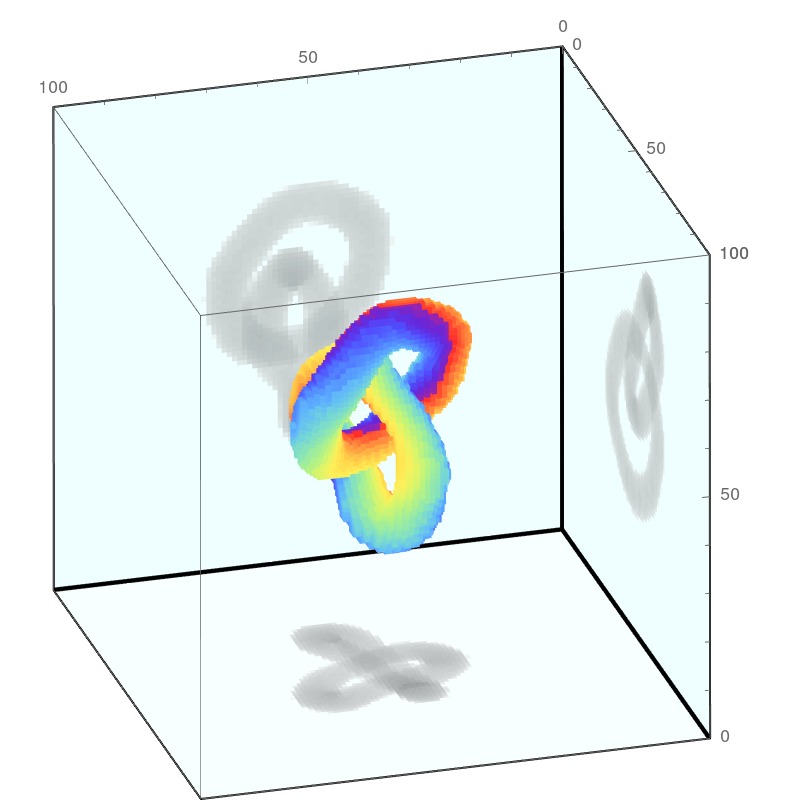}
\par\end{centering}

}\subfigure[$\tau>1.2\times10^{6}$, $\alpha=0.8$, $R=4$]{\begin{centering}
\includegraphics[width=0.32\columnwidth]{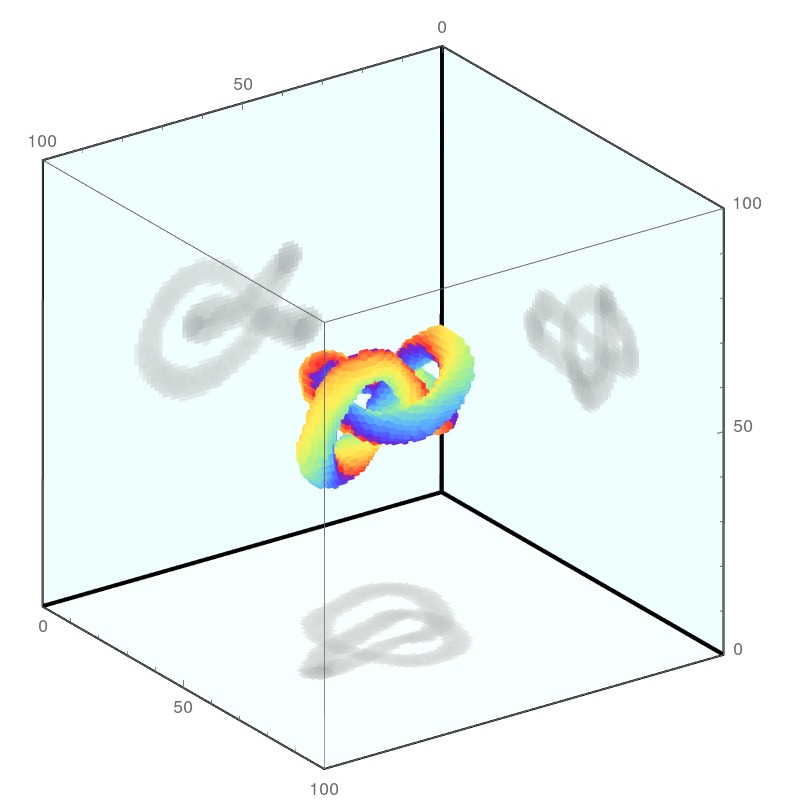}
\par\end{centering}

}\subfigure[$\tau>250000$, $\alpha=0.8$, $R=4$]{\begin{centering}
\includegraphics[width=0.32\columnwidth]{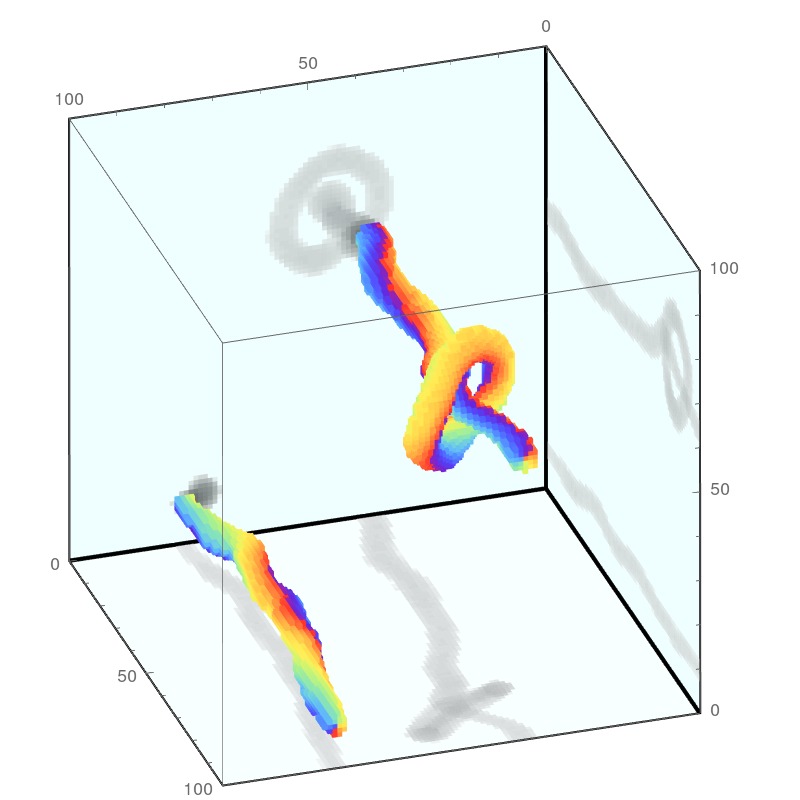}
\par\end{centering}

}
\par\end{centering}

\begin{centering}
\subfigure[$\tau>200000$, $\alpha=0.7$, $R=4$]{\begin{centering}
\includegraphics[width=0.32\columnwidth]{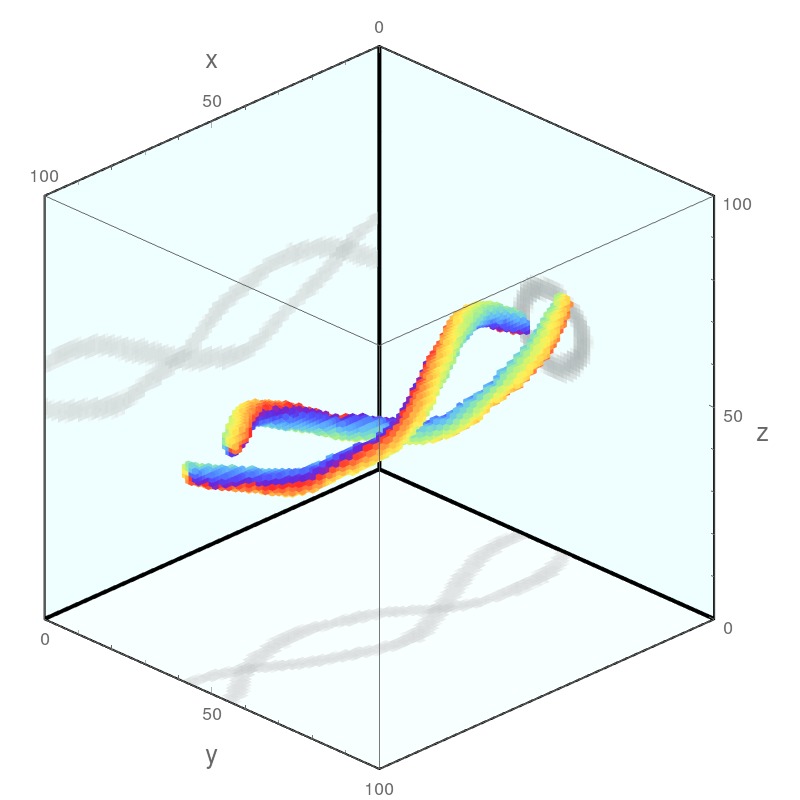}
\par\end{centering}

}\subfigure[$\tau>200000$, $\alpha=0.7$, $R=4$]{\begin{centering}
\includegraphics[width=0.32\columnwidth]{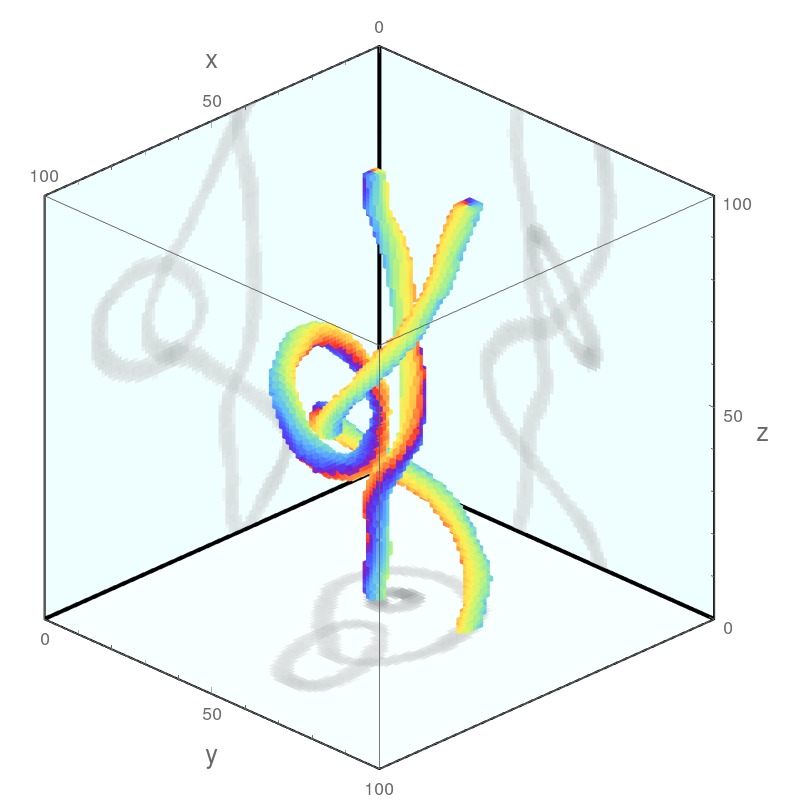} 
\par\end{centering}

}\subfigure[$\tau\sim70000$, $\alpha=0.8$, $R=4$]{\begin{centering}
\includegraphics[width=0.32\columnwidth]{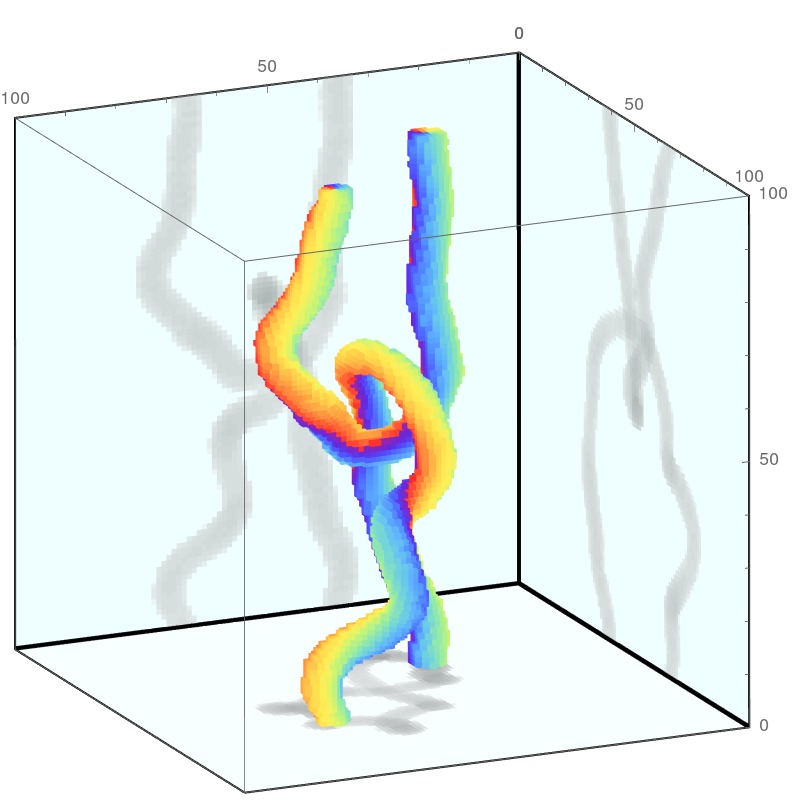}
\par\end{centering}

}
\par\end{centering}

\begin{centering}
\subfigure[$\tau\sim5000$, $\alpha=0.8$, $R=4$]{\begin{centering}
\includegraphics[width=0.32\columnwidth]{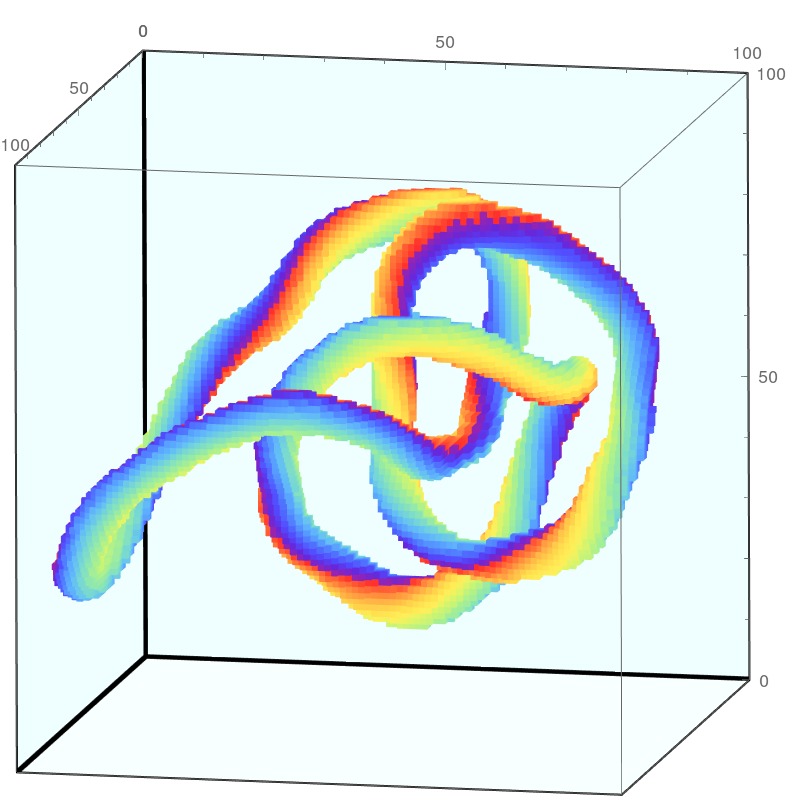}
\par\end{centering}

}\subfigure[$\tau\sim5000$, $\alpha=0.8$, $R=4$]{\begin{centering}
\includegraphics[width=0.32\columnwidth]{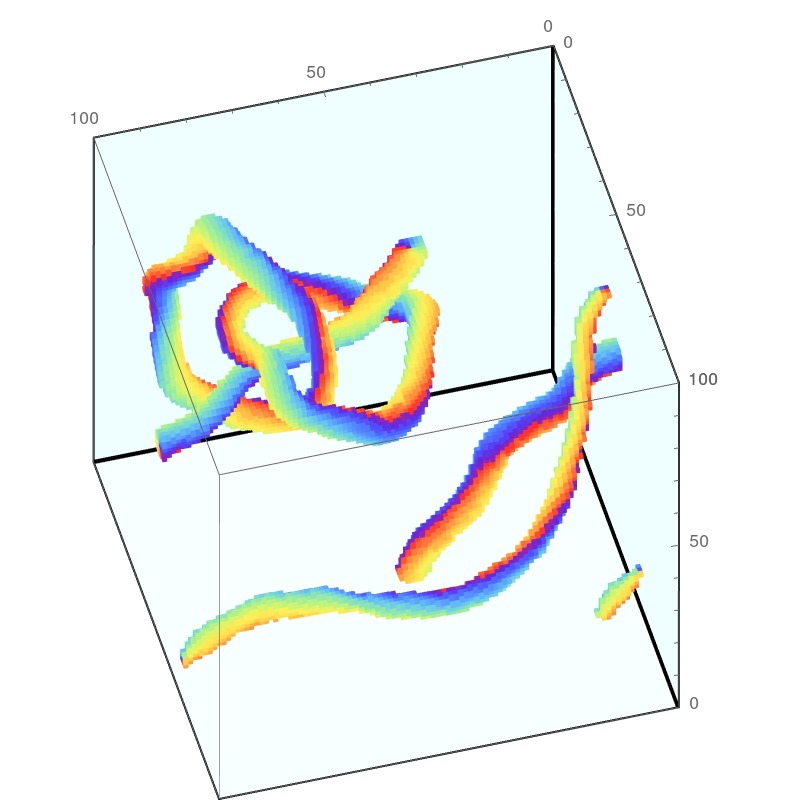}
\par\end{centering}

}\subfigure[$\tau\sim5000$, $\alpha=0.7$, $R=5$]{\begin{centering}
\includegraphics[width=0.32\columnwidth]{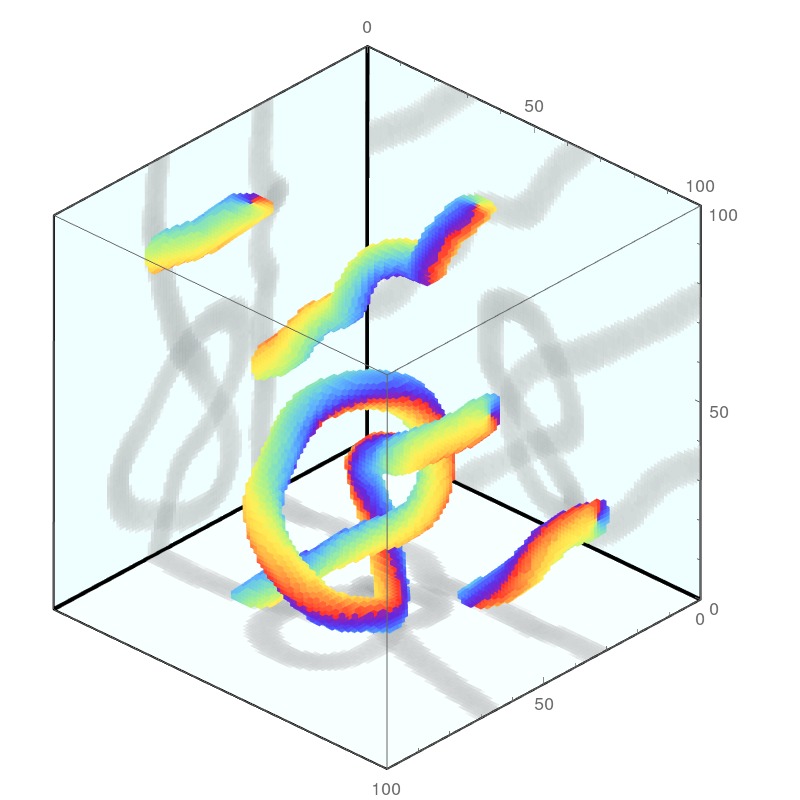}
\par\end{centering}

}
\par\end{centering}

\caption{\label{cap: Topo-struct} Non-trivial topological structures in the non-local
Kuramoto model with periodic BC for $L=100$. The lifetime $\tau$
and the corresponding parameters are given for each subfigure. $\tau>\tau_{0}$
means that the structure is stable within the testing time limit $\tau_{0}$,
while $\tau\sim\tau_{0}$ means the structure breaks down around $\tau_{0}$
(order of magnitude). The period of the scroll waves is about $T\sim11$ for
$\alpha=0.8$. This implies a lifetime of more than $10^{5}T$ for
Hopf links and trefoils. Together with the robustness in the presence of noise as established in the
main text, this suggests that the lifetime $\tau\to\infty$ when $L\gg R$.}
\end{figure*}

\begin{figure*}[p]
\begin{centering}
\subfigure[Two simple filaments with no twist ($\alpha=0.8$).]{\begin{centering}
\includegraphics[width=0.3\columnwidth]{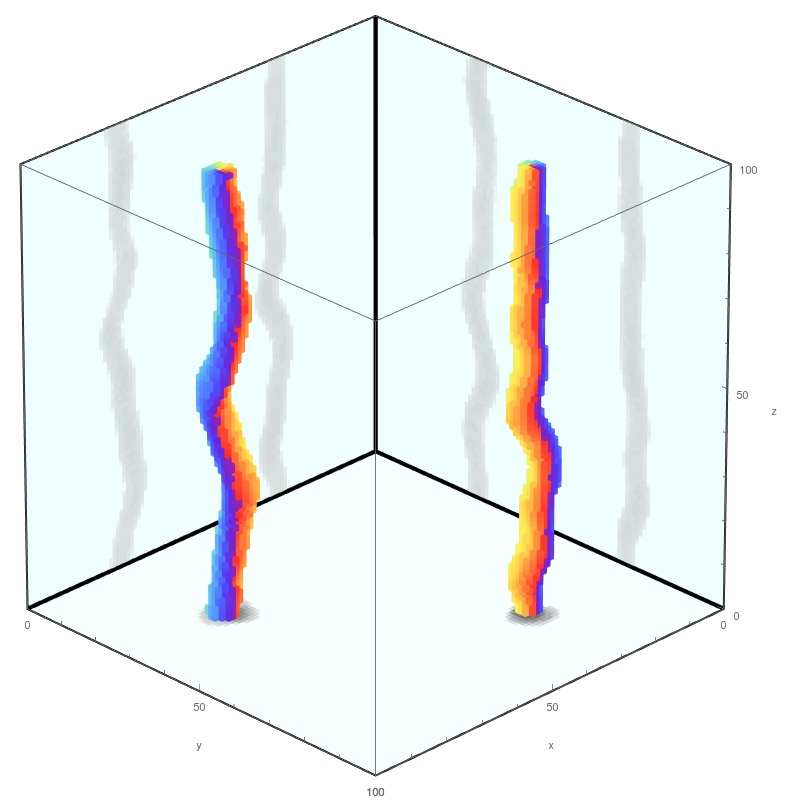}
\par\end{centering}

}\subfigure[Two simple filaments twisting once ($\alpha=0.8$).]{\begin{centering}
\includegraphics[width=0.3\columnwidth]{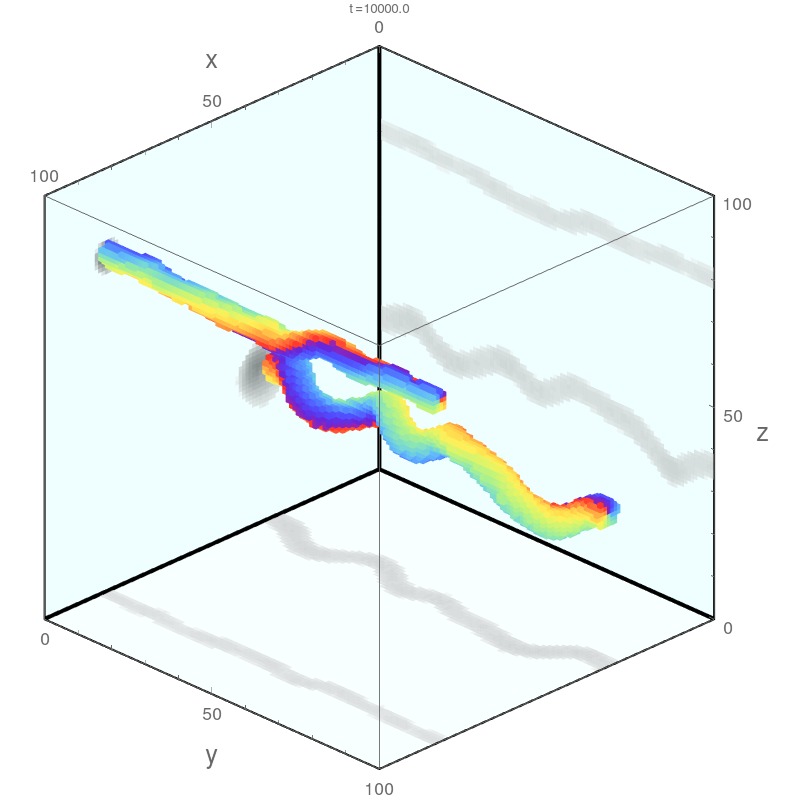}
\par\end{centering}

}\subfigure[Two twisted filaments passing through all three surfaces ($\alpha=0.05$). ]{\begin{centering}
\includegraphics[width=0.3\columnwidth]{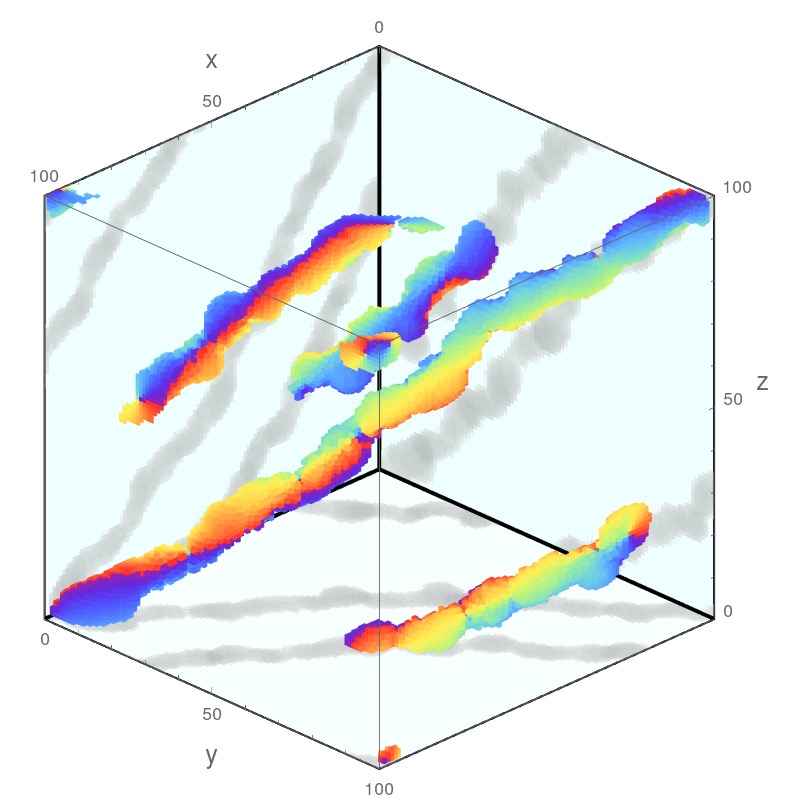}
\par\end{centering}

}
\par\end{centering}

\caption{\label{cap: Tubes} Various long-lived filaments with periodic BC, $L=100$ and $R=4$.
(a-b) Each filament connects with itself through one of the surfaces. (c)
Each filament passes through all three surfaces before connecting back to
itself.}
\end{figure*}

Fig.~\ref{cap: Topo-struct} shows various long-lived (stable or metastable) topological structures in the non-local Kuramoto model within the regime $\alpha_{K}<\alpha<\alpha_{0}$. Note that knotted structures more complicated than simple Hopf links tend to have smaller stability regimes. An exception are (knotted) structures that require periodic boundary conditions (BC) \emph{and} do not drift, which can be stable below $\alpha_{K}$. This is shown, for example, in Fig.~\ref{cap: Tubes} and includes straight filaments. Simulations suggest that the multi-filament structure in Fig.~\ref{cap: Tubes}c is stable for $\alpha>0$.

\section{Dynamics}

\begin{figure*}[p]
\begin{centering}
\subfigure[Hopf link]{\begin{centering}
\includegraphics[width=0.25\columnwidth]{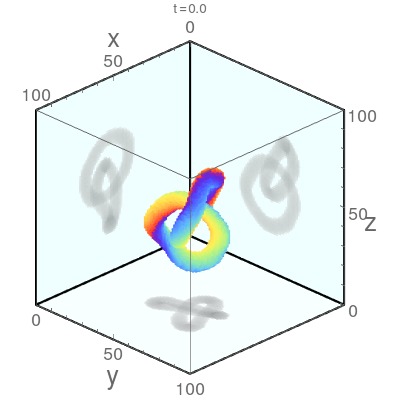}\includegraphics[width=0.25\columnwidth]{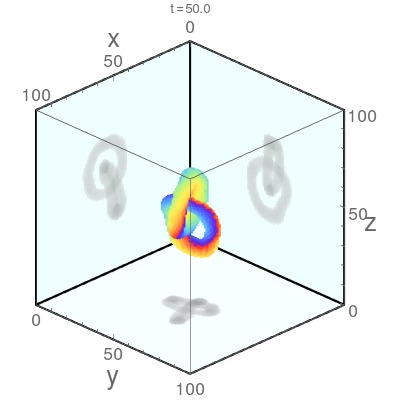}\includegraphics[width=0.25\columnwidth]{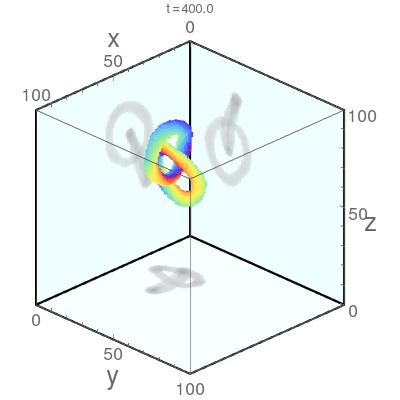}\includegraphics[width=0.25\columnwidth]{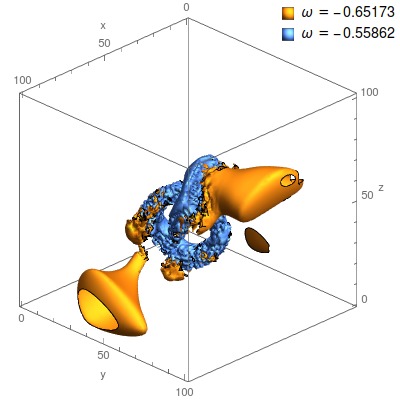}
\par\end{centering}

}
\par\end{centering}

\begin{centering}
\subfigure[Trefoil]{\begin{centering}
\includegraphics[width=0.25\columnwidth]{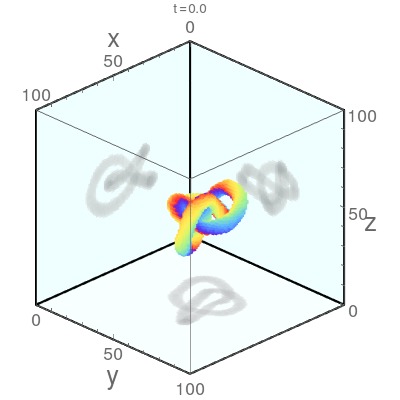}\includegraphics[width=0.25\columnwidth]{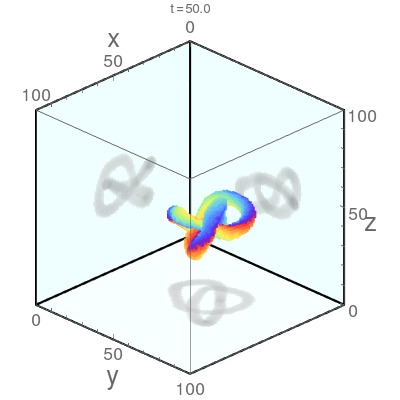}\includegraphics[width=0.25\columnwidth]{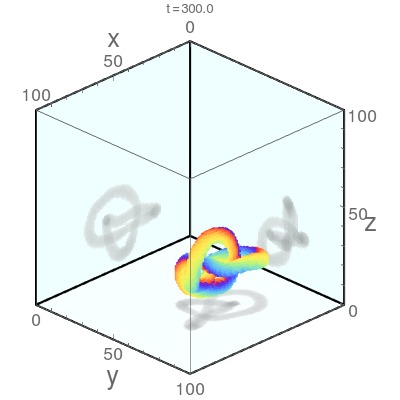}\includegraphics[width=0.25\columnwidth]{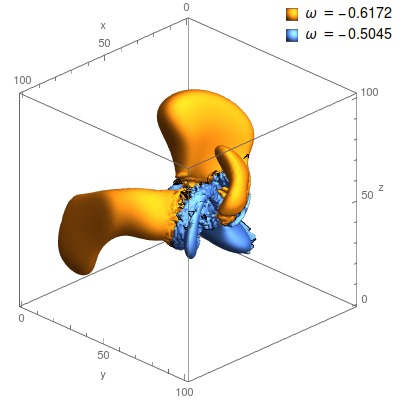}
\par\end{centering}

}
\par\end{centering}

\begin{centering}
\subfigure[Ring-tube]{\begin{centering}
\includegraphics[width=0.25\columnwidth]{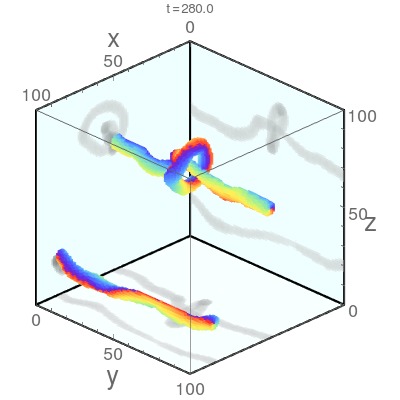}\includegraphics[width=0.25\columnwidth]{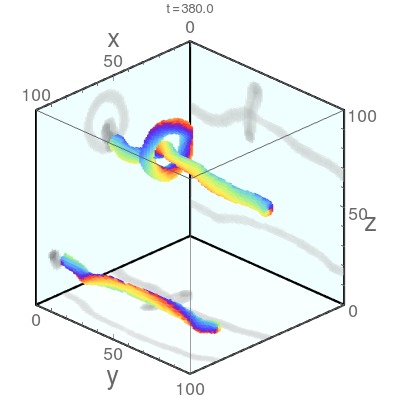}\includegraphics[width=0.25\columnwidth]{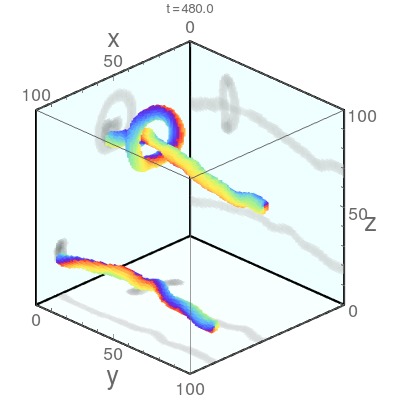}\includegraphics[width=0.25\columnwidth]{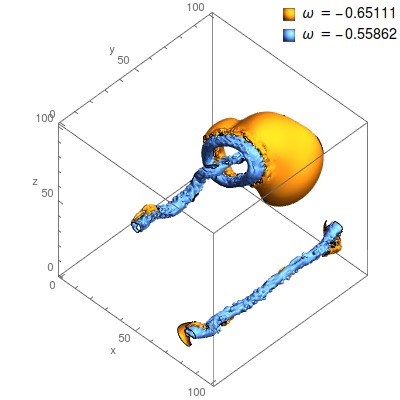}
\par\end{centering}

}
\par\end{centering}

\caption{\label{cap: dynamics} Dynamics of different topological structures ($\alpha=0.8$, $R=4$ and periodic BC).
The first three columns are snapshots at three different instances in time. 
The rightmost column is the iso-surface plot of the instantaneous
angular frequency $\omega$ of the last snapshot. Blue
indicates the region with $|\omega|<|\bar{\omega}|$, while
orange indicates the region $|\omega|>|\bar{\omega}|$.
Filaments are moving away from the orange region.}
\end{figure*}

In the non-local Kuramoto model, knotted structures that exist independent of the specific choice of BC (periodic vs. no-flux) are not stationary but drift, rotate and change their shape over time. As an example, Figs.~\ref{cap: dynamics}(a) and ~\ref{cap: dynamics}(b) show a few snapshots for different structures. Note that for the system sizes studied, the center of mass motion is not straight over long time scales. The phase field away from these knotted structures takes on the form of spherical waves as shown in Fig. \ref{cap: rr-spherical-wave}.
In case of the ring-tube structure (which is specific to periodic BC), the ring propagates along the tube and keeps distorting the local part of the tube while it travels, see Fig.~\ref{cap: dynamics} (c). In all these cases, the direction of the filament motion can be deduced from the instantaneous angular frequency $\omega(x,y,z)$ shown in the rightmost column of Fig.~\ref{cap: dynamics}. 

\begin{flushleft}
\begin{figure*}[p]
\begin{centering}
\includegraphics[width=0.4\columnwidth]{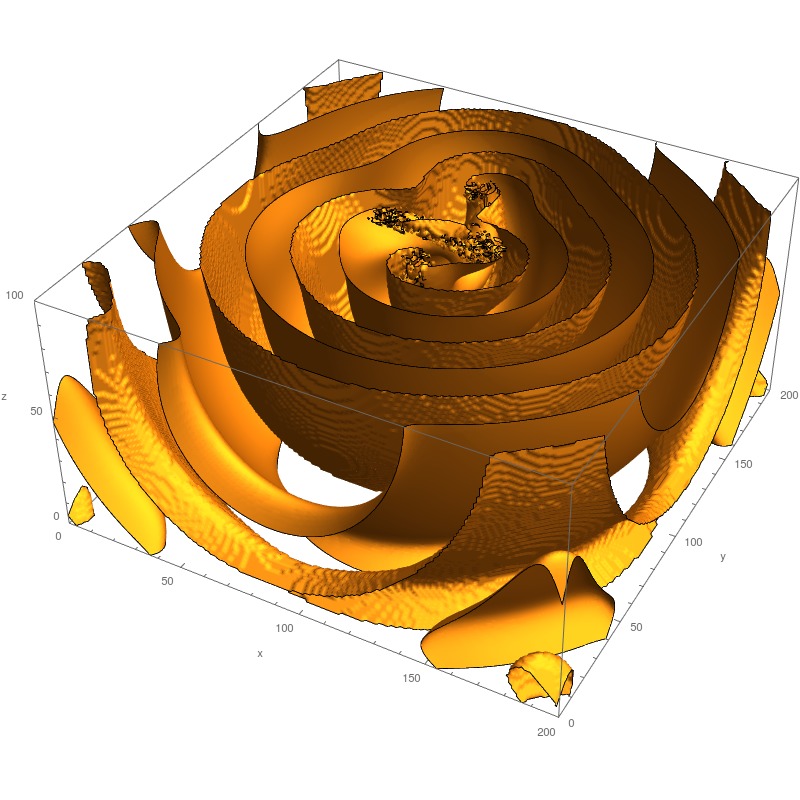}
\par\end{centering}

\caption{\label{cap: rr-spherical-wave} The spherical wave generated around
a Hopf link is shown by plotting the iso-surface $\theta=0$ ($\alpha=0.8$, $R=8$ and periodic BC). The
irregular pattern at the center is the region with unsynchronized
oscillators that form the filament. Note that only the lower half of the system is shown to highlight the structures near the center.}
\end{figure*}

\par\end{flushleft}

\section{Creating chimera knots}

\subsection{Random initial condition}

\begin{figure*}[p]
\begin{centering}
\subfigure[A transient state for $R=8$, $L=200$.]{\begin{centering}
\includegraphics[width=0.32\columnwidth]{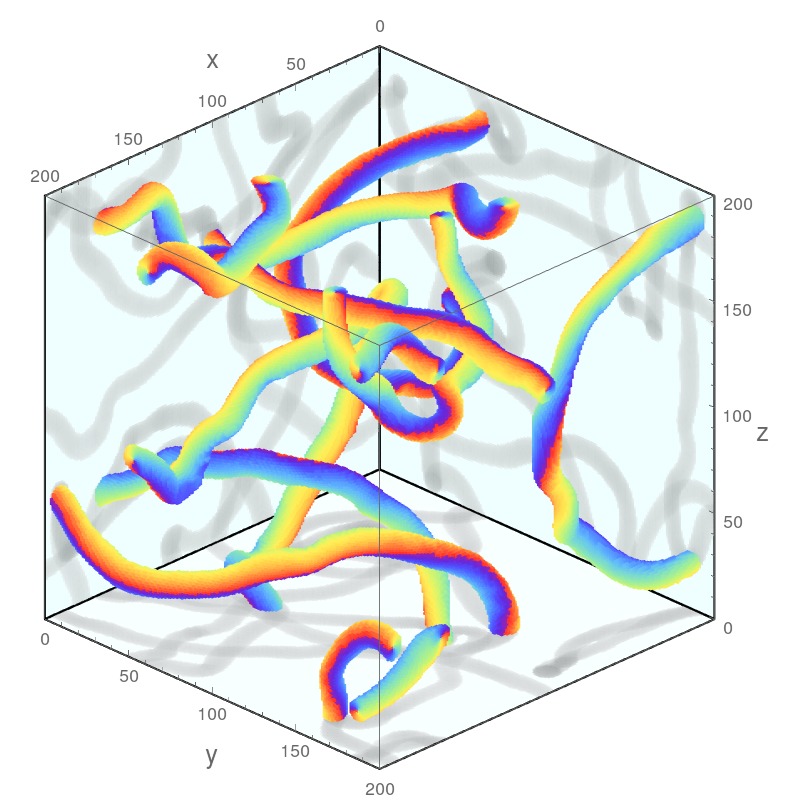}
\par\end{centering}

}\subfigure[A transient state for $R=4$, $L=200$. See Sec.~\ref{sec:reconnecting} for a discussion of the significance of the red box.]{\begin{centering}
\includegraphics[width=0.32\columnwidth]{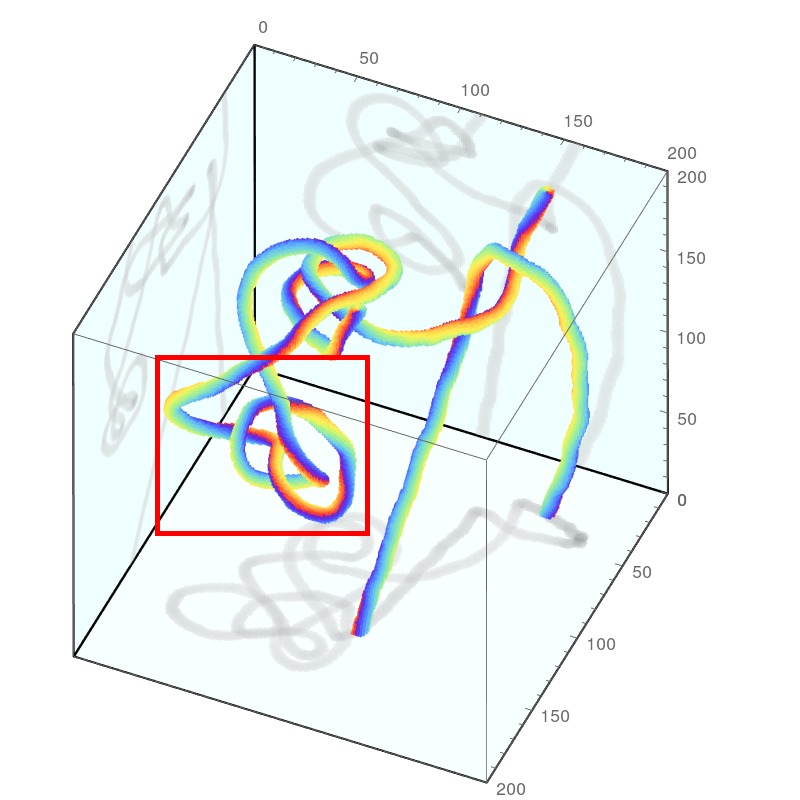}
\par\end{centering}

}\subfigure[Same as in (b) at a later time.]{\begin{centering}
\includegraphics[width=0.32\columnwidth]{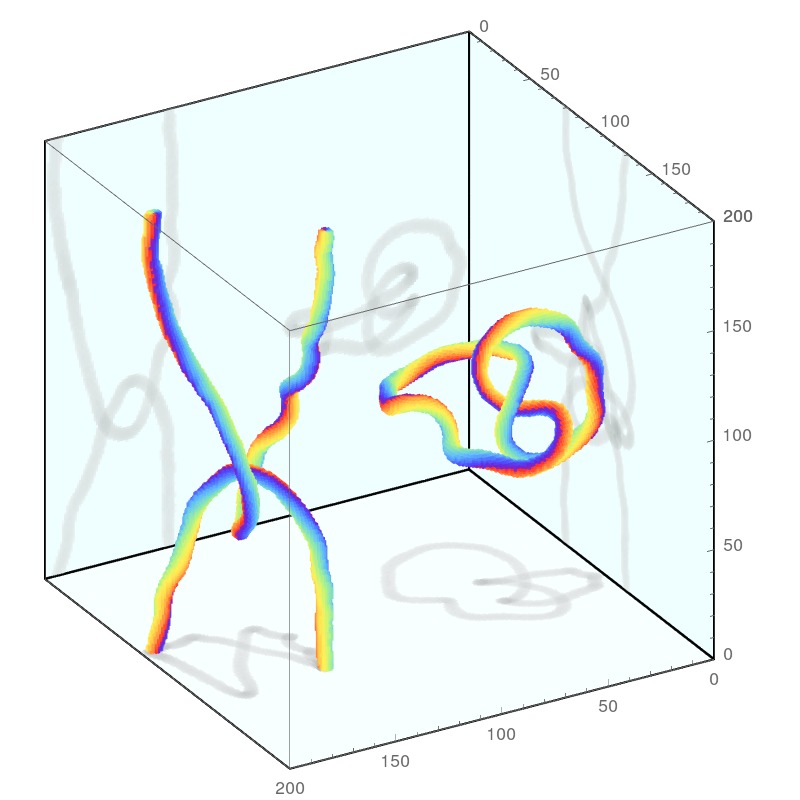}
\par\end{centering}

}
\par\end{centering}

\caption{\label{cap: eg-random-ic} Some snapshots of transient states ($\alpha=0.8$ and periodic BC).}
\end{figure*}

Knots and links can appear spontaneously from random initial conditions (IC). The transient time is of the order of one thousand scroll wave periods in the regimes being studied. A few snapshots of typical transient states are shown in Fig.~\ref{cap: eg-random-ic}. Using random IC, we can obtain all knotted structures shown in Figs.~\ref{cap: Topo-struct}a-\ref{cap: Topo-struct}f. The specific probabilities of generating Hopf links and trefoils from random IC are summarized in Table \ref{tab: struct-occurence}.

\begin{table}[p]
\caption{\label{tab: struct-occurence} Spontaneous formation of Hopf
links and trefoils in simulations with random IC and periodic BC.}

\centering{}%
\begin{tabular}{|c|c|c|c|c|c|}
\hline 
$\alpha$ & $R$ & $L$ & Number of simulations & Number of Hopf links & Number of trefoils\tabularnewline
\hline 
\hline 
0.7 & 4 & 100 & 200 & 1 & 1\tabularnewline
\hline 
0.7 & 5 & 100 & 500 & 4 & 0\tabularnewline
\hline 
0.8 & 4 & 100 & 500 & 13 & 0\tabularnewline
\hline 
0.8 & 5 & 100 & 500 & 5 & 0\tabularnewline
\hline 
0.7 & 5 & 200 & 100 & 3 & 1\tabularnewline
\hline 
0.8 & 8 & 200 & 100 & 4 & 0\tabularnewline
\hline 
\end{tabular}
\end{table}

\subsection{Algorithm to create rings and Hopf links \label{sec:ic}}

\begin{figure*}[p]
\begin{centering}
(a)\includegraphics[width=0.3\columnwidth]{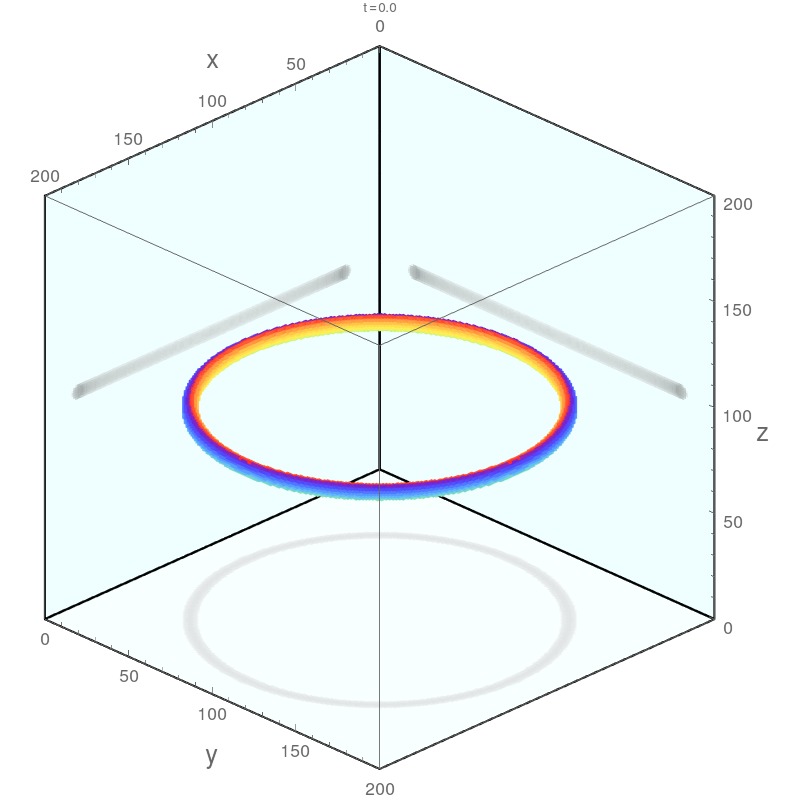}\includegraphics[width=0.3\columnwidth]{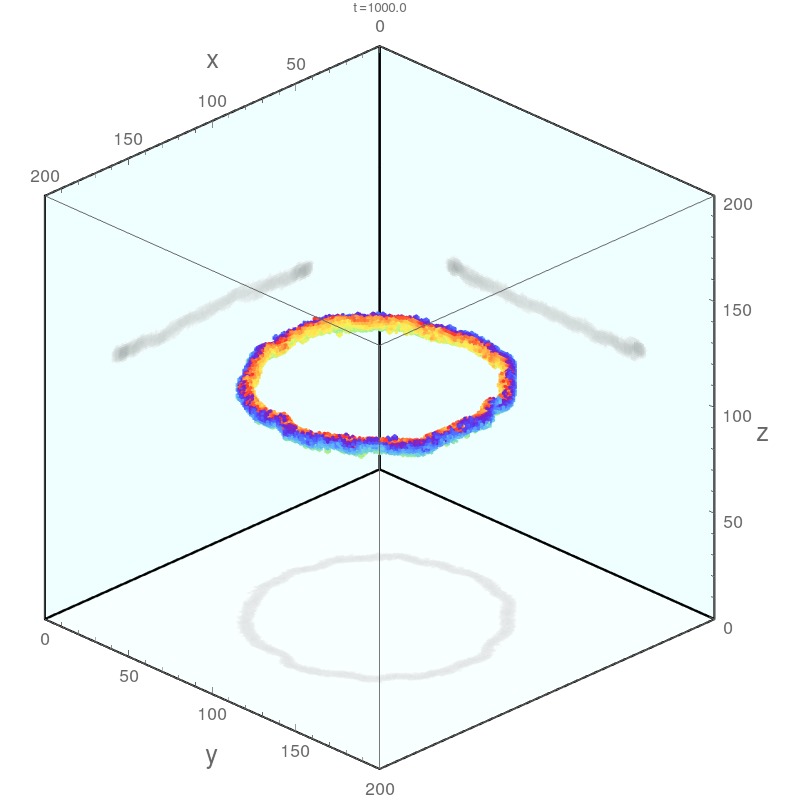}\includegraphics[width=0.3\columnwidth]{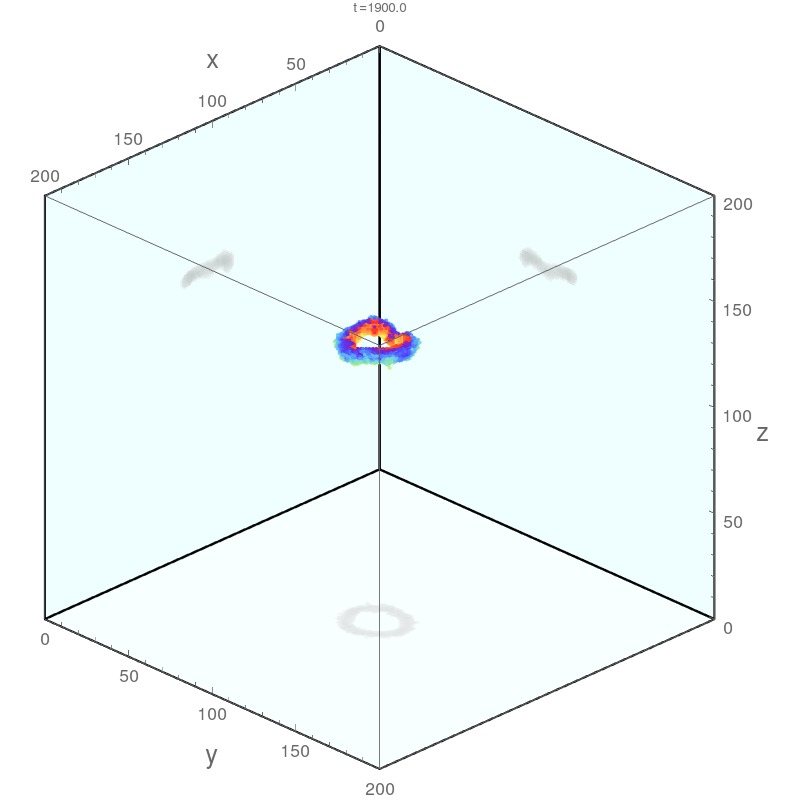}
\par\end{centering}

\begin{centering}
(b)\includegraphics[width=0.3\columnwidth]{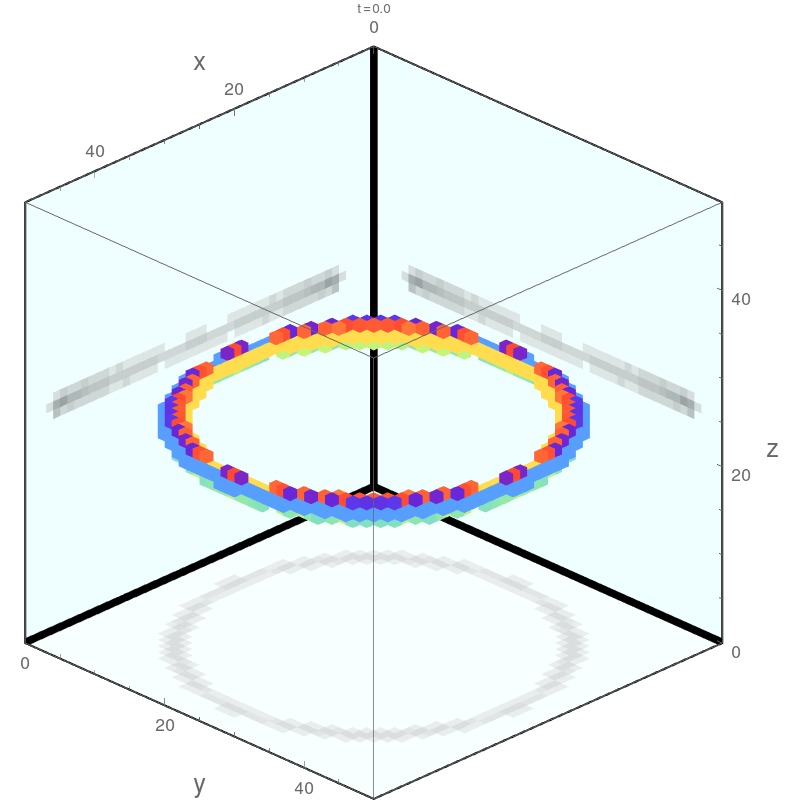}\includegraphics[width=0.3\columnwidth]{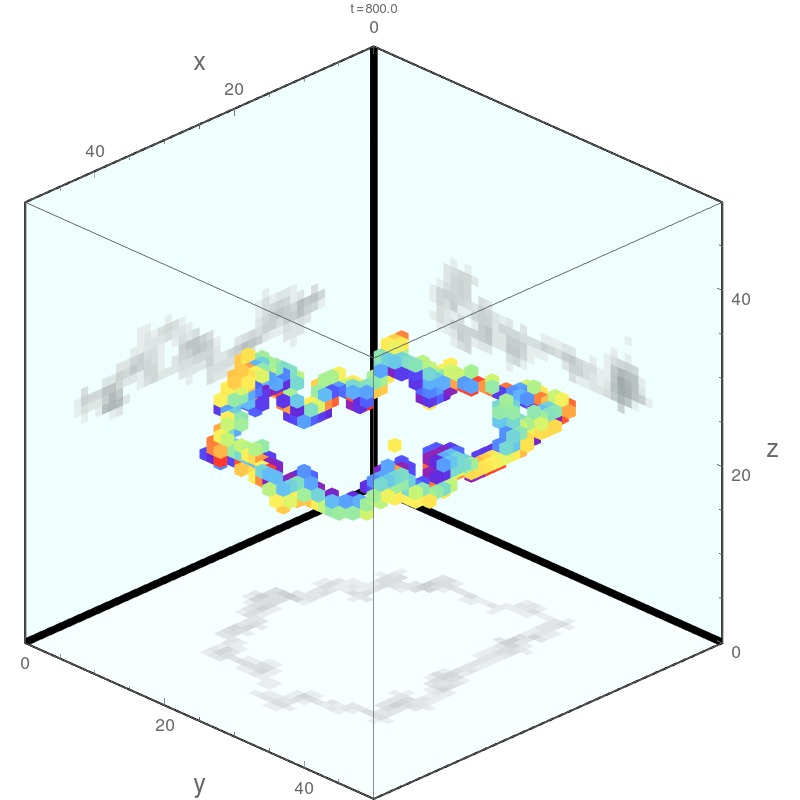}\includegraphics[width=0.3\columnwidth]{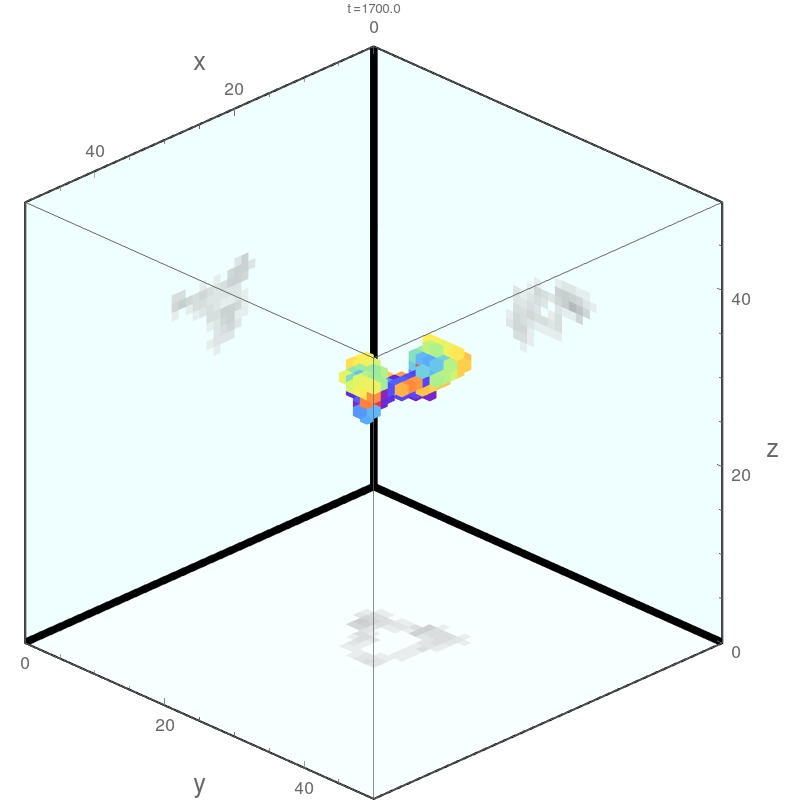}
\par\end{centering}

\caption{\label{cap: shrinking-ring} Shrinking rings with (a) non-local coupling
$R=4$, (b) nearest-neighbor coupling $R=1$. Parameters: $L/R=50$,
$R_{0}/R=20$, $\alpha=0.8$ with no-flux BC.}
\end{figure*}

First, the phase field of a single ring is considered. Suppose the
center of a ring is located at $\mathbf{r}_{0}=(x_{0},y_{0},z_{0})$
with radius $R_{0}$ and the normal vector of the ring is pointing
in the positive $\hat{z}$ direction. A parameterization of the location
of this ring using $\phi\in[0,2\pi)$ is
\begin{equation}
\mathbf{r}(\phi)=(r_{x},r_{y},r_{z})=(x_{o}+R_{0}\cos\phi,y_{0}+R_{0}\sin\phi,z_{0}).
\end{equation}
To create a ring shaped filament corresponding to phase singularities, the phase field
needs to be specified in the whole domain such that it is smooth outside
the ring but results in $2\pi$ phase difference while going around
a point on the filament. This can be done by defining
\begin{eqnarray}
\varphi & = & \tan^{-1}\left(\frac{z-z_{0}}{R_{0}-f}\right)\\
f & = & \sqrt{(x-x_{0})^{2}+(y-y_{0})^{2}}
\end{eqnarray}
for any spatial point $\mathbf{r}=(x,y,z)$. Then the phase of each oscillator
$\theta(\mathbf{r})$ can be computed by $\theta(\mathbf{r})=\psi(\mathbf{r})$,
where
\begin{equation}
\psi(\mathbf{r})=kd-\varphi-s\phi-\beta.
\end{equation}
Here, $k$ is the wavenumber, $d$ is the distance to the closest
point on the ring $d=\sqrt{(x-r_{x})^{2}+(y-r_{y})^{2}+(z-r_{z})^{2}}$,
$\varphi$ is the angle between the plane consisting of the ring and
the line to the closest point of the ring, $s$ is the twisting number,
$\phi$ is the ring parameterization, and $\beta$ is a constant phase
shift. Examples are shown in Fig. \ref{cap: shrinking-ring}.

\begin{figure*}[p]
\begin{centering}
(a)\includegraphics[width=0.32\columnwidth]{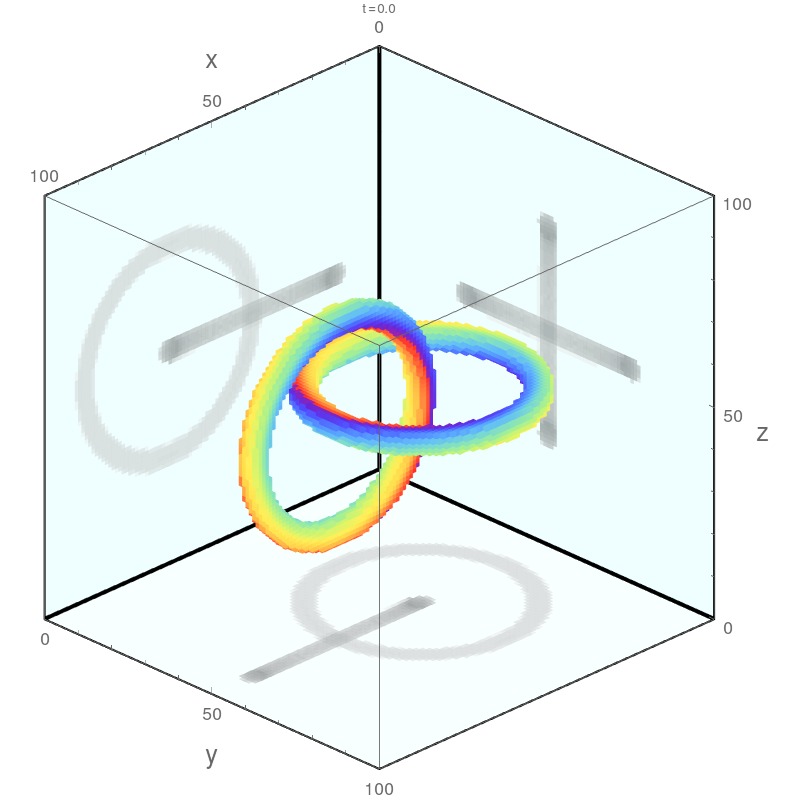}\includegraphics[width=0.32\columnwidth]{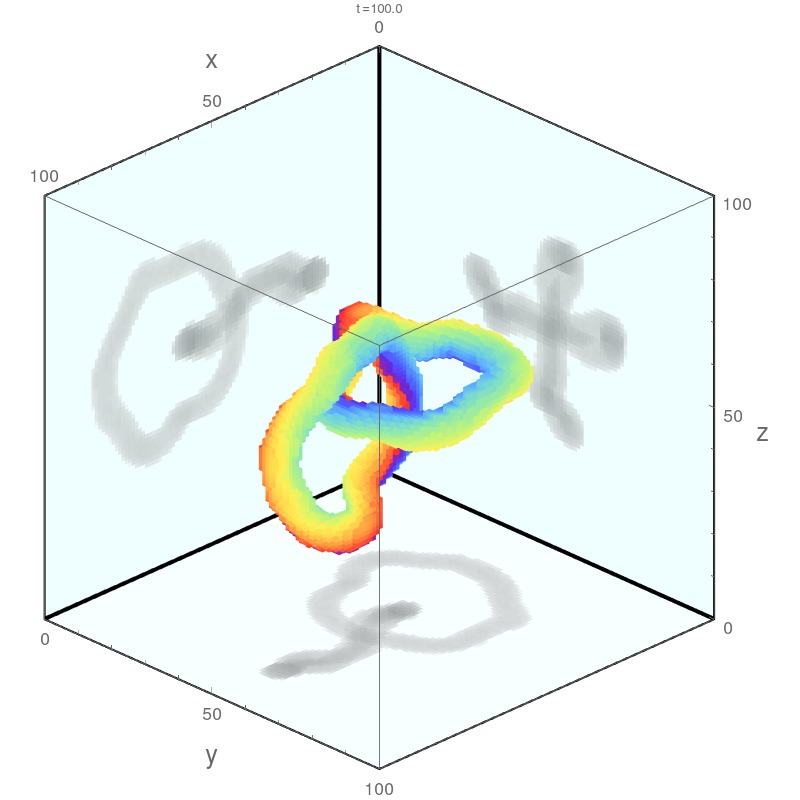}\includegraphics[width=0.32\columnwidth]{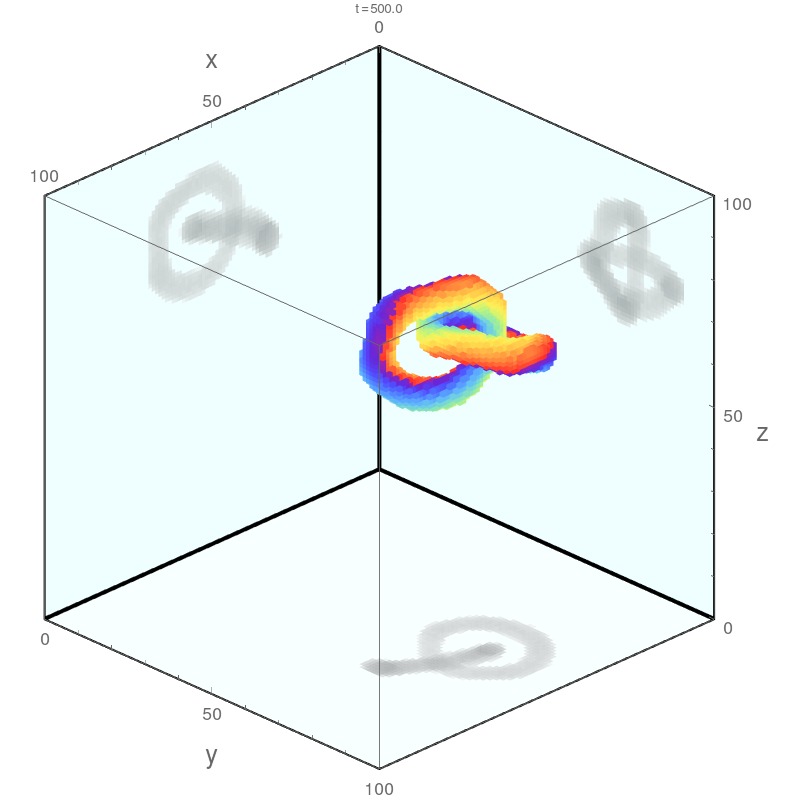}
\par\end{centering}

\begin{centering}
(b)\includegraphics[width=0.32\columnwidth]{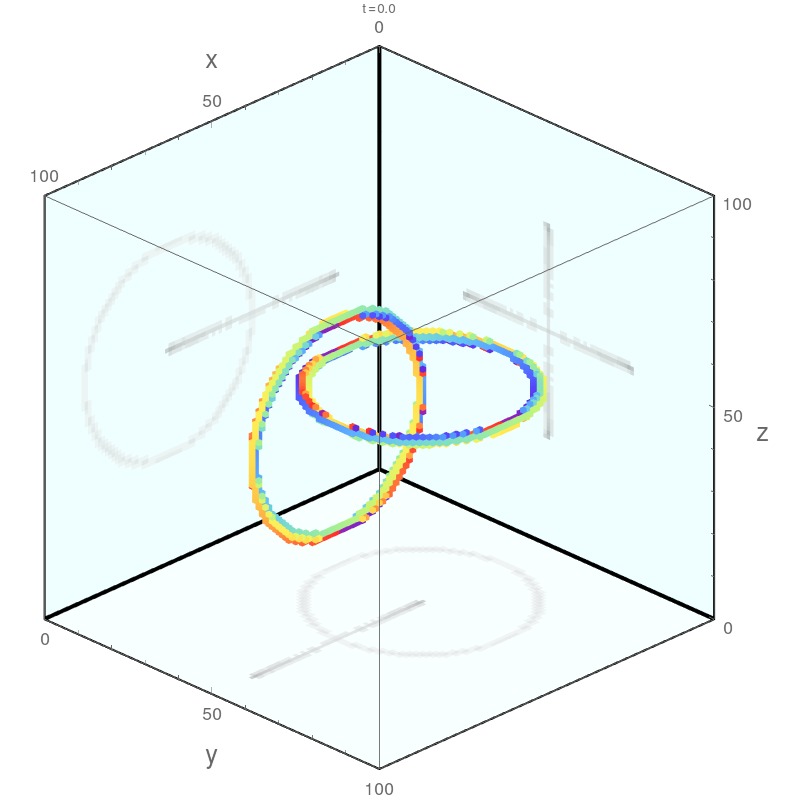}\includegraphics[width=0.32\columnwidth]{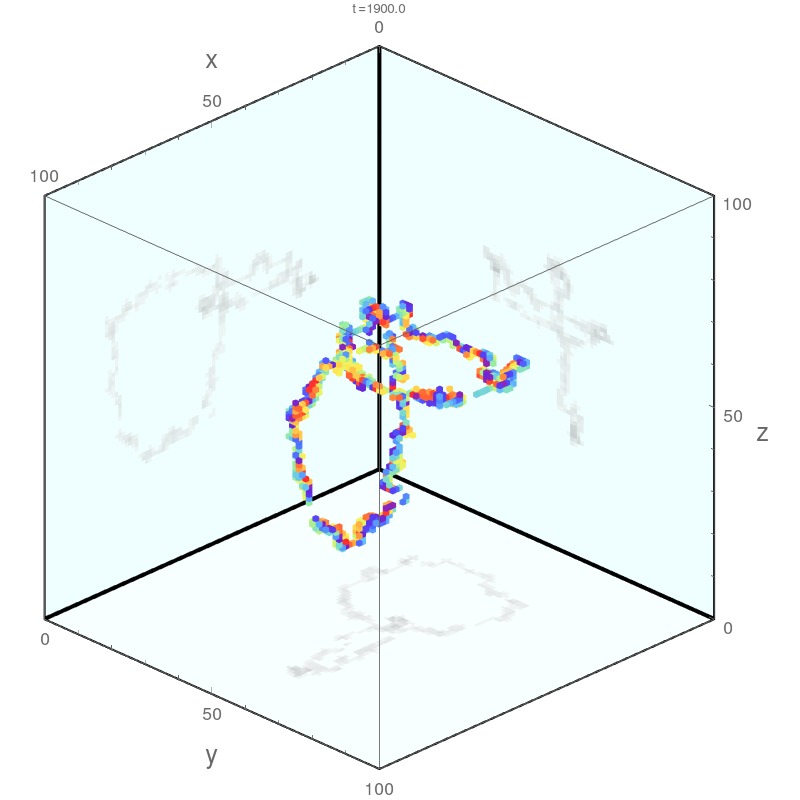}\includegraphics[width=0.32\columnwidth]{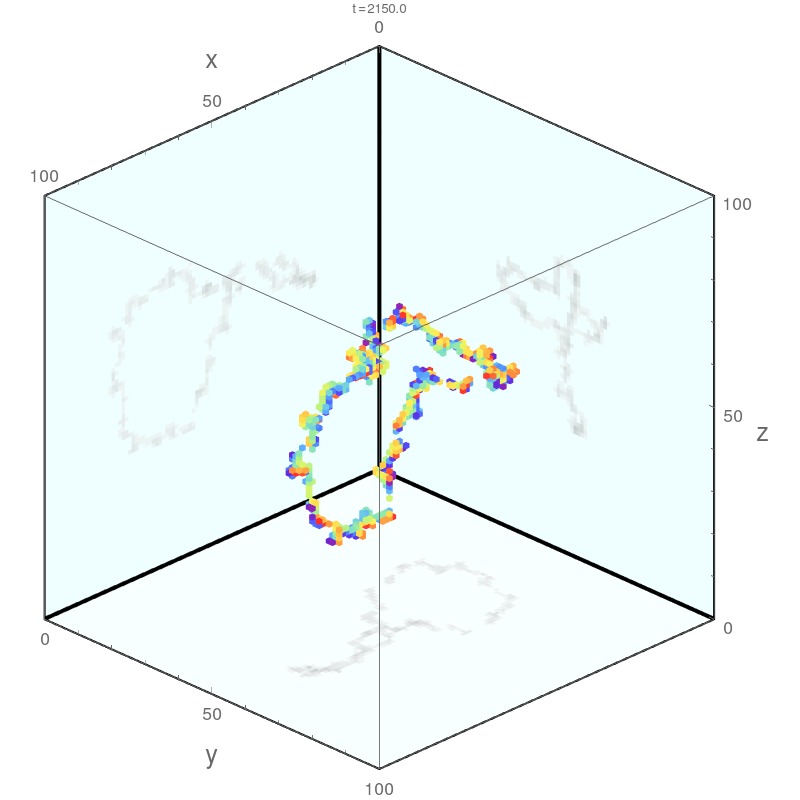}
\par\end{centering}

\caption{\label{cap: shrinking-rr} Formation of knots for (a) non-local coupling
$R=4$, (b) nearest neighbor coupling $R=1$. Note that the IC
are exactly the same in both cases. Parameters: $R_{0}=25$, $L=100$,
$\alpha=0.8$ with no-flux BC.}
\end{figure*}

The phase field of a Hopf link can be created by combining two rings,
requiring a method to smoothly superimpose them. This can be achieved
using a distance dependent phase:
\begin{equation}
\xi(\mathbf{r},\mathbf{r}_{0},s,\beta)=\left(\frac{R}{d}\right)^{2}e^{i\psi(\mathbf{r},s,\beta)},
\end{equation}
which is based on the inverse square distance. Then the phase field
of a Hopf link $\theta(\mathbf{r})$ can be calculated by 
\begin{equation}
\rho(\mathbf{r})e^{i\theta(\mathbf{r})}=\xi(x,y,z,x_{0}-R_{0}/2,y_{0},z_{0},s,\beta=0)+\xi(x,z,y,x_{0}+R_{0}/2,y_{0},z_{0},s,\beta=\pi)
\end{equation}
with twisting number $s=1$.  Examples are shown in Fig. \ref{cap: shrinking-rr}.
Note that a structure in a given system size $L$ can be rescaled to
$L'$ using a simple scaling function of the form $\theta'(x',y',z')=\theta(\left\lfloor \frac{L}{L'}x'\right\rfloor ,\left\lfloor \frac{L}{L'}y'\right\rfloor ,\left\lfloor \frac{L}{L'}z'\right\rfloor )$,
where $\left\lfloor \cdot\right\rfloor $ denotes the floor of the
number (which is necessary since the oscillators are arranged on a discrete lattice) and the prime denotes the new phase and new location. This
rescaling works quite well for the top-hat kernel as long as $R\sim R'\gg1$. Also, if the
smoothed phase $\tilde{\theta}$ of knots --- see Eq.~(4) in the main text --- is used as IC, the unsynchronized
region around the filaments can redevelop.

\subsection{Reconnecting chimera filaments using random patches \label{sec:reconnecting}}

A new structure can be obtained by reconnecting local filaments of
a known structure. This reconnection requires a detailed specification
of the whole local phase field that is smooth, without creating
new filaments and while matching the desired filaments. This can be hard to
do if the local filaments are obtained from a simulation. Alternatively,
based on the observation that only simple straight filament can form
in a small system size $L/R$ from random IC, it suggests a way to
transform a structure by randomizing a whole local region. Using this
method, we have successfully created trefoils and a few other knots.
To begin with, a structure that is similar to the desired knot
is needed, with the region of reconnection close to each other. For
example, the structure in the red box shown in Fig.~\ref{cap: eg-random-ic}
is a trefoil if the top parts are connected. After half a dozen trials
using different shapes of the randomized region, we were indeed able to create a trefoil.
Note that the region should be large enough to form a tube but not
too large to form other structures. This method may suggest a similar
way to create knots in real world experiment.

\section{Instabilities}

\begin{figure*}[p]
\begin{centering}
\includegraphics[width=0.4\columnwidth]{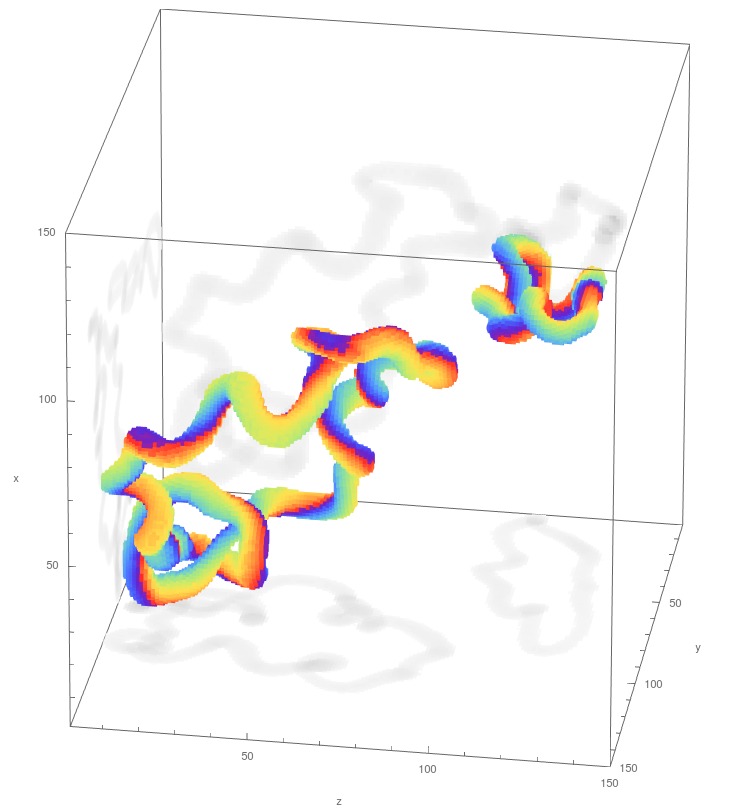}
\par\end{centering}

\caption{\label{cap: trefoil-near-a0} Instability of a trefoil near the transition
point $\alpha_{0}$ ($R=4$, $L=150$ and periodic BC). This snapshot shows the initial elongation 
of one branch of the trefoil, which has collided with itself and formed
an extra ring.}
\end{figure*}

In the main text, the instabilities at $\alpha_{K}$ and $\alpha_{0}$
of Hopf links in the Kuramoto model have been discussed. The instability near $\alpha_{K}$
is caused by a lack of repulsion to counter curvature-driven shrinkage, 
so knots collapse and disappear. On
the other hand, the instability near $\alpha_{0}$ originates from
an instability of the filament where the filaments become longer and longer
and eventually collide with themselves or other filaments. This effect
is particularly clear in large domains as shown, for example, in Fig.~\ref{cap: trefoil-near-a0}.
Note that an elongation also happens as a transient state when the
parameters are suddenly changed or starting from a non-perfect IC.
However, it will eventually shorten after refolding to an asymptotic
state as also observed for other models~\cite{S_winfree}. Other instabilities
are discussed below.

\subsection{Instability of a single ring}

Direct simulations show that rings are not stable for $\alpha<\alpha_{0}$
with no-flux BC. As shown in Fig. \ref{cap: shrinking-ring}, all
rings shrink in size and eventually vanish. The largest ring tested
had radius $R_{0}=80$. This shrinkage process occurs for both nearest
neighbor coupling $R=1$ and non-local coupling $R=4$. 
Note that the time it takes for a ring to disappear is approximately the same in both cases for the same effective radius $R_{0}/R$ and effective system size $L/R$. Also, almost all transient (knotted) states resulting eventually in homogeneous oscillations become rings in their penultimate stage.

\subsection{Instability of knots for $R=1$}

As shown in Fig.~\ref{cap: shrinking-rr}, using the IC for Hopf links described in Section~\ref{sec:ic}
can result in a stable knot if $R\gg1$. For the choice of $R_0$, the two rings initially shrink
in size and then an effective repulsion prevents further shrinkage.
At the same time, the center of the Hopf link starts moving. In contrast,
if the same IC is used with nearest neighbor coupling $R=1$, the two
rings will eventually collide with each other and decay into a single
ring, which in turn shrinks and vanishes.

\subsection{Filament instability at $\alpha_{0}$}

\begin{figure*}[p]
\begin{centering}
(a)\includegraphics[width=0.32\columnwidth]{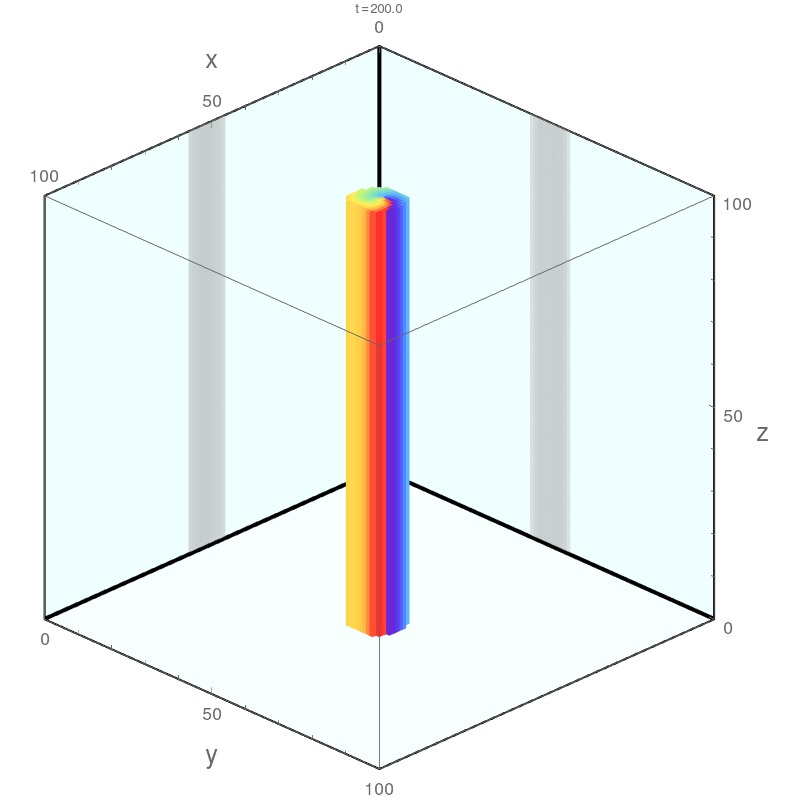}\includegraphics[width=0.32\columnwidth]{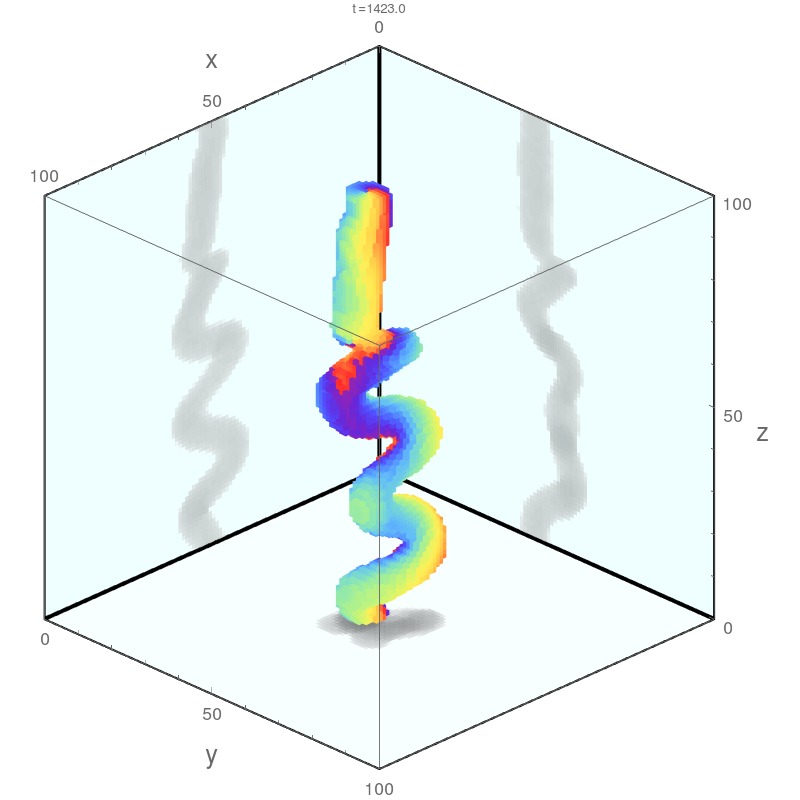}\includegraphics[width=0.32\columnwidth]{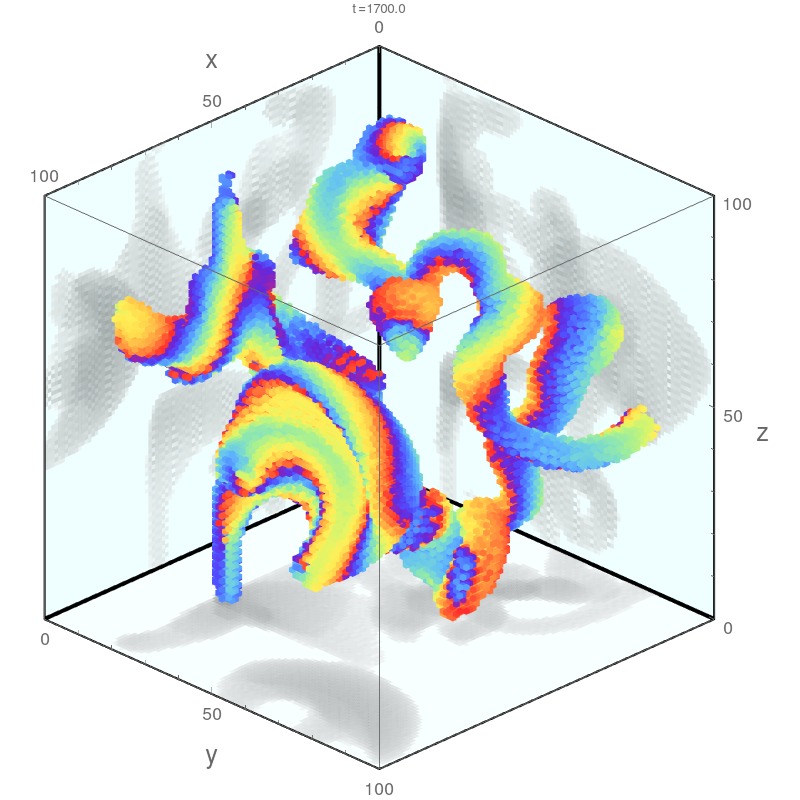}
\par\end{centering}

\begin{centering}
(b)\includegraphics[width=0.32\columnwidth]{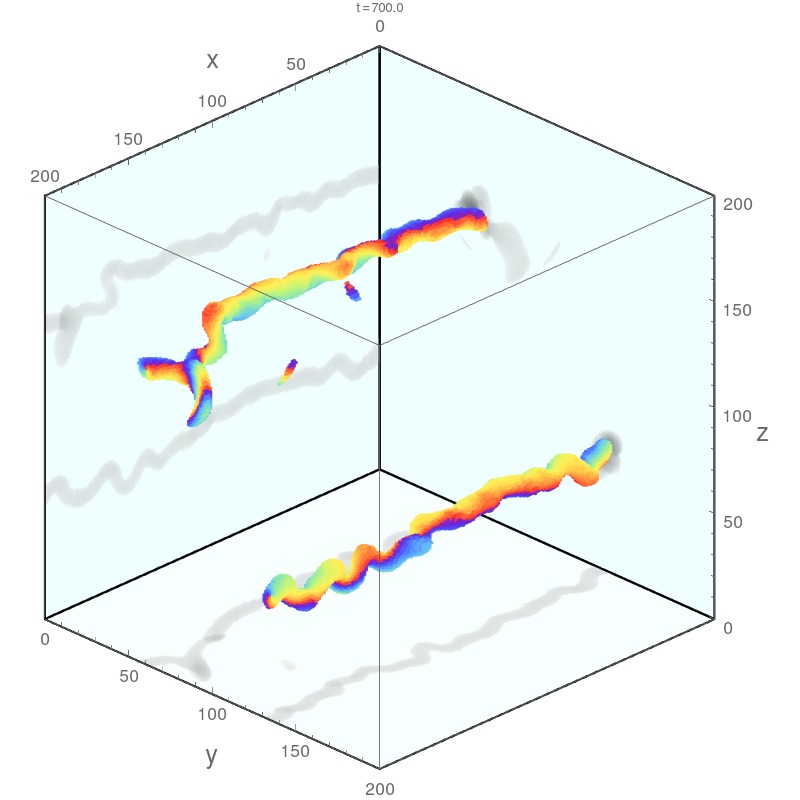}\includegraphics[width=0.32\columnwidth]{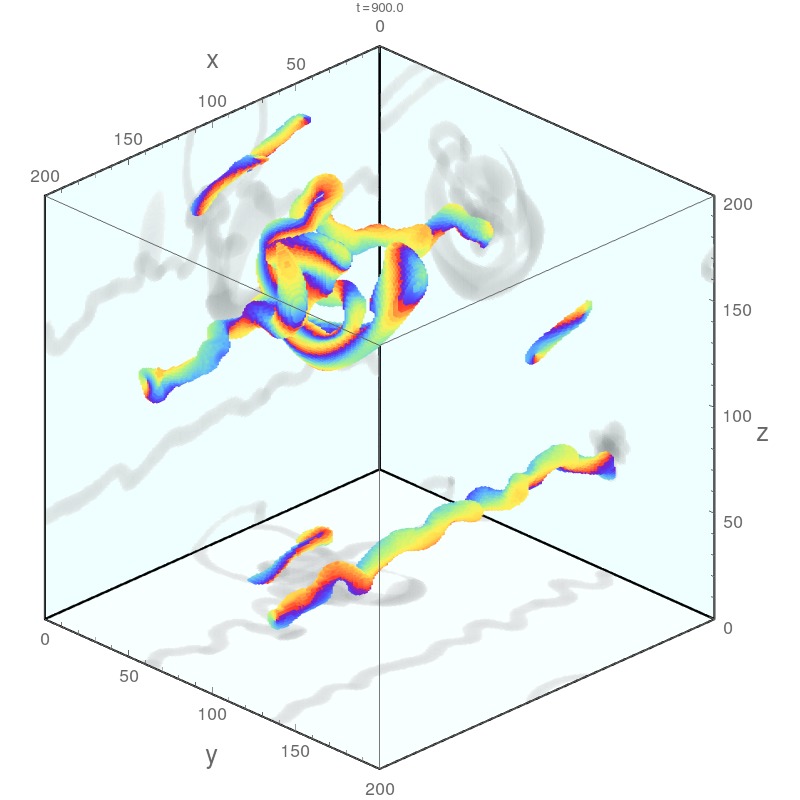}\includegraphics[width=0.32\columnwidth]{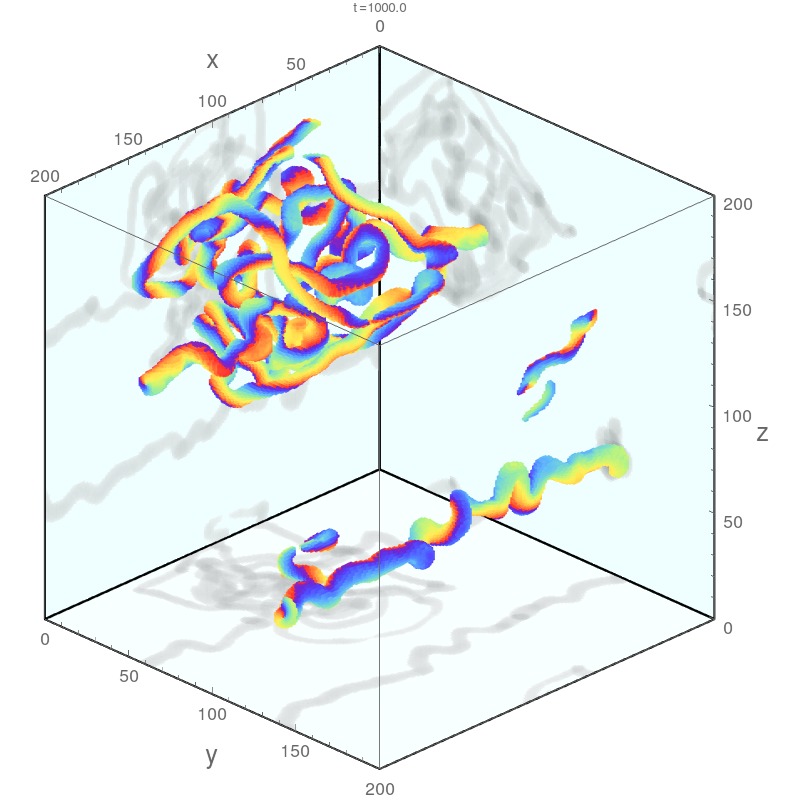}
\par\end{centering}

\caption{\label{cap: tt-alpha0-instability} Snapshot series of the instability of
straight filaments at $\alpha=0.95>\alpha_{0}$. (a) Single filament
with $L=100$, $R=4$ and no-flux boundary conditions. Some secondary structures
develop with local twisting before break-up. (b) Two filaments
in a larger domain $L=200$ and $R=4$ with PBC. The
rapid elongation of one of the filaments is evident.}
\end{figure*}

As illustrated in Fig. \ref{cap: tt-alpha0-instability}, the instability at $\alpha_{0}$ for simple straight chimera filaments is characterized by the emergence of secondary structures and the elongation of filaments. The same qualitative behavior is observed for knotted structures in the regime $\alpha>\alpha_{0}$ as shown in the main text. Nevertheless, the knotted structures can persist for thousands of scroll wave periods before they break up consistent with critical slowing down near a phase transition.

\subsection{Instabilities from finite size effects}

\begin{figure*}[p]
\begin{centering}
\includegraphics[width=0.32\columnwidth]{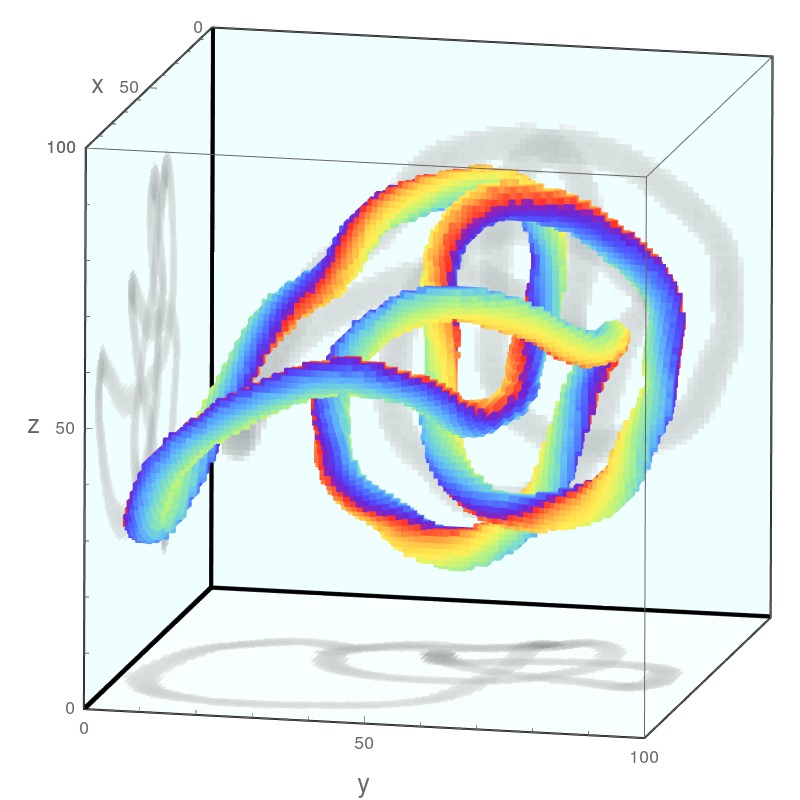}\includegraphics[width=0.32\columnwidth]{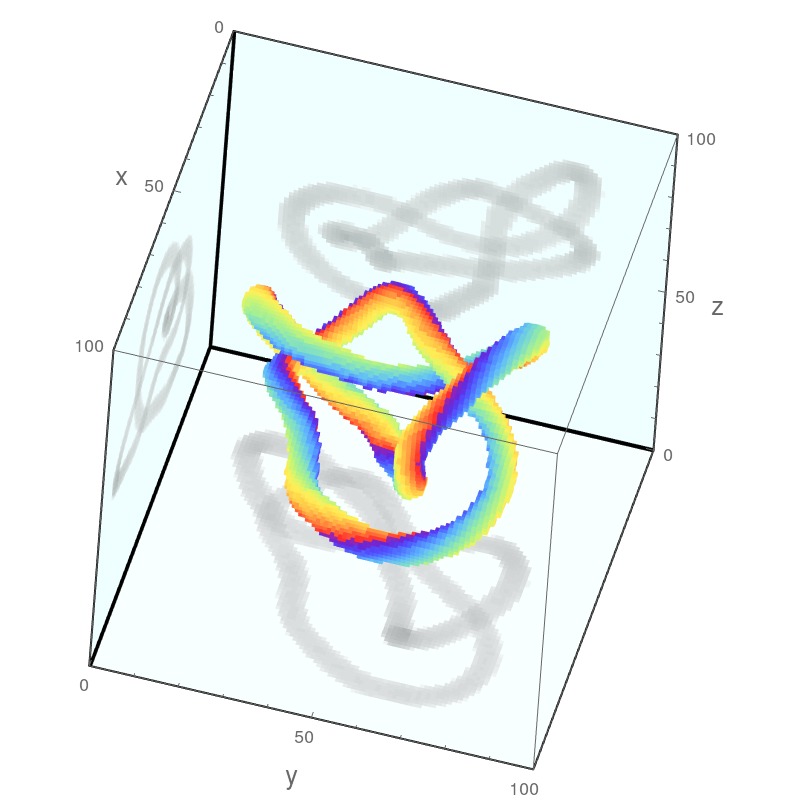}\includegraphics[width=0.32\columnwidth]{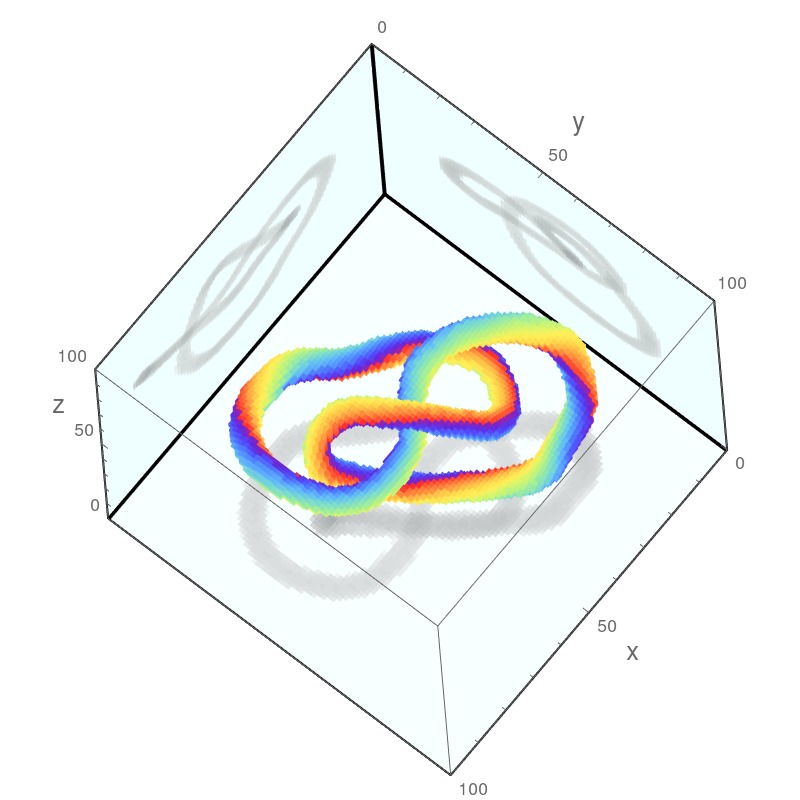}
\par\end{centering}

\caption{\label{cap: rrr-decay} Decay of a triple ring for $L=100$, $R=5$, $\alpha=0.7$ with PBC.
(a) $t=0$, triple rings. (b) $t=15000$, decay into a ring knotted
with an 8-shape ring. (d) $t=20000$, further decay into a trefoil.}
\end{figure*}

A stable knotted structure becomes unstable when it is confined in a small effective system $L/R$. While a Hopf link simply decays into a single ring which eventually vanishes, the situation is more complicated for larger and more complex knotted structures. One example is shown in Fig.~\ref{cap: rrr-decay} starting from a triple ring for $R=4$ in $L=100$, which decays into a ring knotted with 8-shape ring, and then transforms into a trefoil. Depending on the IC and the exact parameter regime, the decay path can be different. Note that even for the moderately larger system size $L=150$, triple rings have significantly longer lifetimes ($\tau>20000$) in some parameter regimes.

\section{Spatial kernels}

\begin{figure*}[p]
\begin{centering}
\includegraphics[width=0.4\columnwidth]{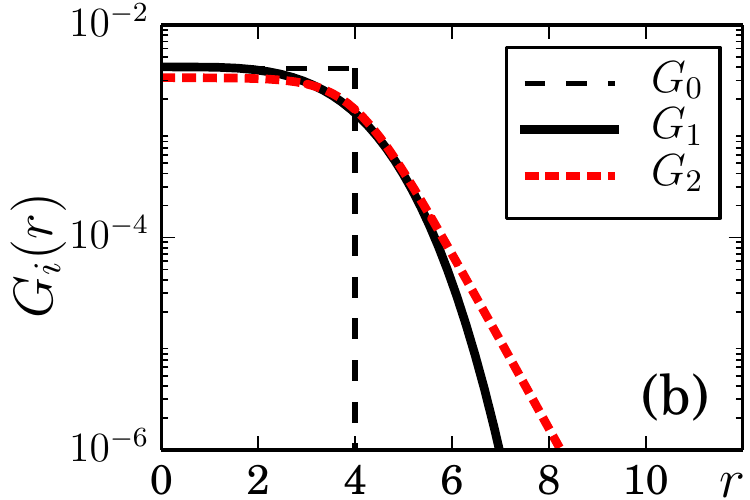}
\par\end{centering}

\caption{\label{cap: kernels}(color online) Plot of the localized kernels that can still result in a stable Hopf link, with estimated critical values $n_{1}=3.8$ and $n_{2}=1.9$ for $R=4$ and $\alpha=0.8$ of Kuramoto model. $G_0$ is shown for comparison.
}
\end{figure*}

As mentioned in the main text, our main findings do not depend qualitatively on the exact functional form of the considered kernels. 
%Specifically, knots can remain stable across various kernels. 
For example, using the kernel
\begin{equation}
G_{0}'(\mathbf{r})\sim\begin{cases}
1, & |x|,|y|,|z|\le R\\
0, & \mbox{otherwise}
\end{cases}
\end{equation}
instead of the top-hat kernel $G_0$ gives pretty much identical results for the stability of knots. 
As another example, using the kernel
\begin{equation}
G_{2}(\textbf{r}) \sim (1+e^{n_{2}(r-R)})^{-1}
\end{equation}
with an exponential tail instead of the kernel $G_1$ with super-exponential tail exhibits the same phenomenology: With decreasing $n_2$, the stable regime of knots shrinks. The shape of the kernels at the transition points of $G_1$ and $G_2$ for $R=4$ and $\alpha=0.8$ are shown in Fig.~\ref{cap: kernels}.

\section{Other oscillatory models}

\subsection{Non-Local Complex Ginzburg-Landau equation (CGLE)}

\begin{figure*}[p]
\begin{centering}
\subfigure[$K=0.1$]{\begin{centering}
\includegraphics[width=0.4\columnwidth]{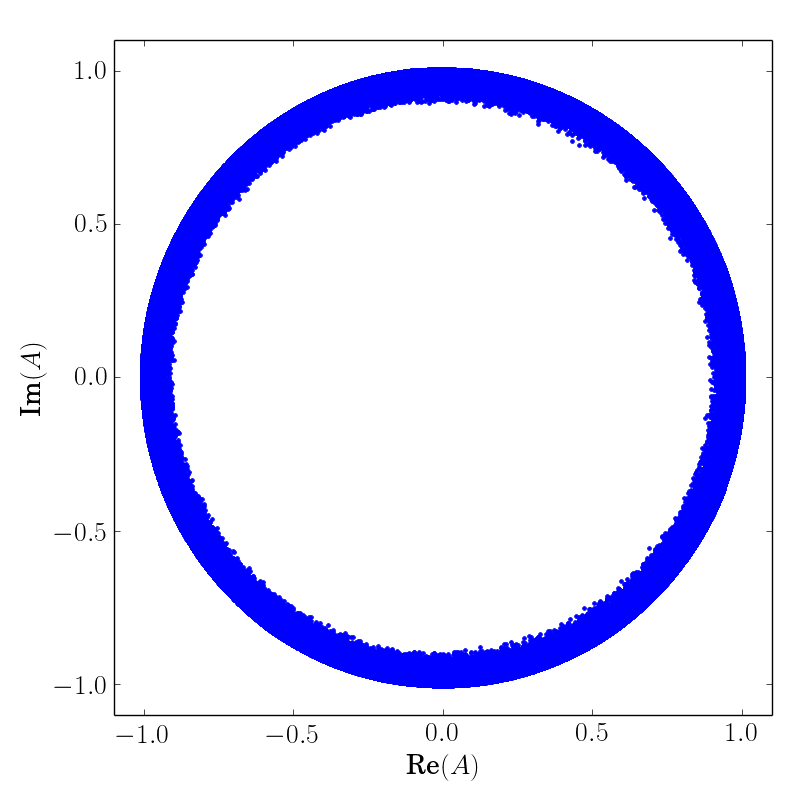}
\par\end{centering}

}\subfigure[$K=0.2$]{\begin{centering}
\includegraphics[width=0.4\columnwidth]{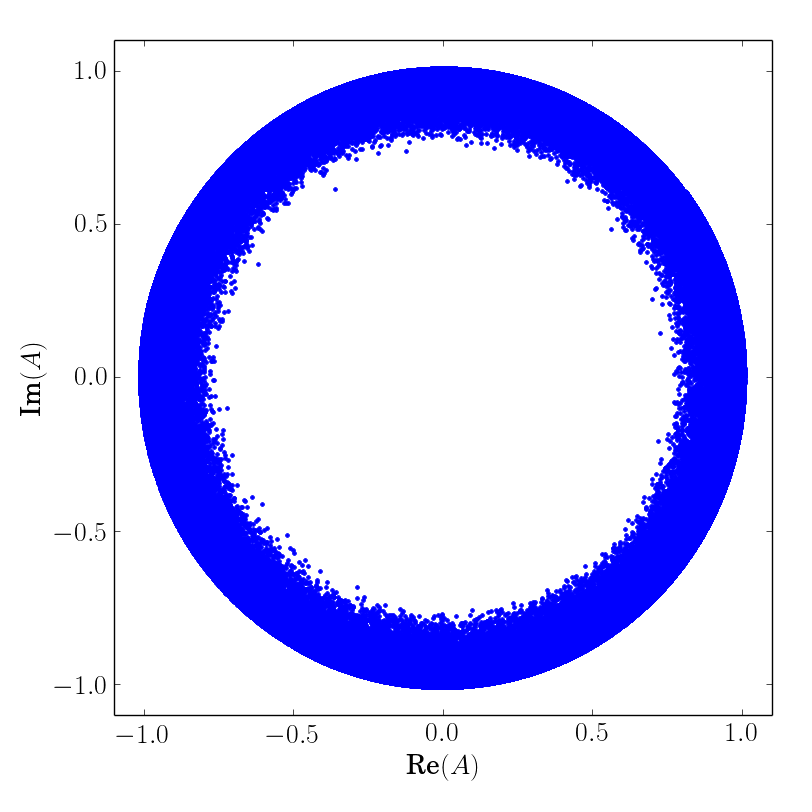}
\par\end{centering}

}
\par\end{centering}

\caption{\label{cap: cgle-phase-portrait} 
Snapshot of the states of the oscillators in phase space for a Hopf link in the non-local CGLE for $(a,b)=(1,0)$, $L=200$, $R=8$ and periodic BC.
}
\end{figure*}

\begin{figure*}[p]
\begin{centering}
\subfigure[Hopf link]{\begin{centering}
\includegraphics[width=0.4\columnwidth]{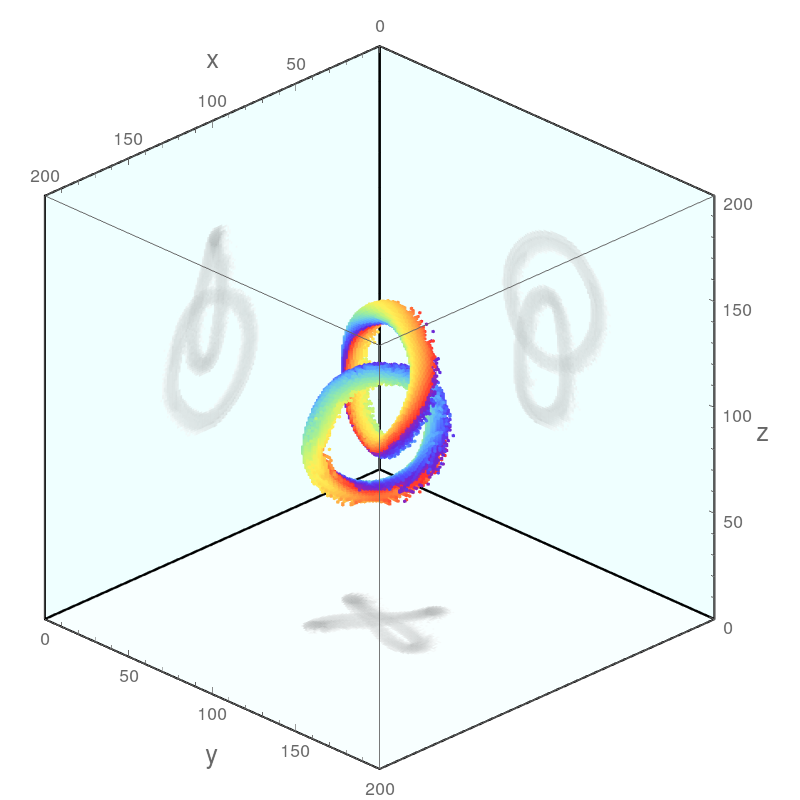}
\par\end{centering}

}
\par\end{centering}

\begin{centering}
\subfigure[$\mbox{Re}(A)$]{\begin{centering}
\includegraphics[width=0.33\columnwidth]{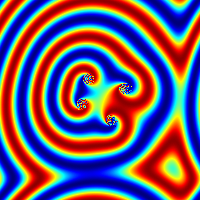}
\par\end{centering}

}\subfigure[$\mbox{Im}(A)$]{\begin{centering}
\includegraphics[width=0.33\columnwidth]{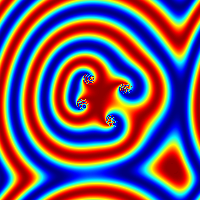}
\par\end{centering}

}\subfigure[$\theta=\arg(A)$]{\begin{centering}
\includegraphics[width=0.33\columnwidth]{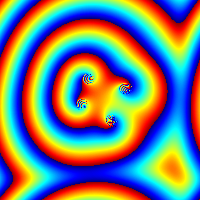}
\par\end{centering}

} 
\par\end{centering}

\caption{\label{cap: cgle-knot} Snapshot of a Hopf link in the non-local CGLE corresponding to Fig.~\ref{cap: cgle-phase-portrait}(a). (a) shows the unsynchronized region corresponding to the chimera knot. An x-y cross-section of the different fields at $z=100$ is plotted in (b)-(d). In (b) and (c), the color map from deep blue to red corresponds to values from $-1$ to $1$ in the respective field.
}
\end{figure*}

The non-local CGLE considered here is \cite{S_kuramoto}:
\begin{equation}
\dot{A}(\mathbf{r},t)=A-(1+ib)|A|^{2}A+K(1+ia)\int G(\mathbf{r}-\mathbf{r'})(A(\mathbf{r}')-A(\mathbf{r}))d\mathbf{r}',
\end{equation}
where the control parameters are $(a,b)$, the coupling strength is $K$ and $G=G_0$ in the following.
Under sufficiently weak coupling $K\to0$, the local field oscillates
with unit amplitude $|A|\approx1$ and behaves like a simple phase
oscillator in the non-local Kuramoto model. Therefore, we can use the knotted structures found in 
the Kuramoto model as IC by simply setting $A(\mathbf{r},t=0)=e^{i\theta(\mathbf{r})}$.
We find that one of the regimes with stable Hopf links is $0.95\lesssim a\lesssim1.15$
for $b=0$ and $K=0.1$ provided that $L \gg R \gg 1$. In Fig.~\ref{cap: cgle-phase-portrait}(a), the phase
portrait shows that the magnitude of all oscillators only deviates slightly
from $|A|=1$ in this case.
For stronger coupling $K=0.2$, the deviations in $A$ increase (see Fig.~\ref{cap: cgle-phase-portrait}(b)) but stable knots still exist.
In both cases, the phase $\theta(\mathbf{r})=\arg(A(\mathbf{r}))$ behaves similar to the Kuramoto model as confirmed by Fig.~\ref{cap: cgle-knot}(d). As Figs.~\ref{cap: cgle-knot}(b) and~\ref{cap: cgle-knot}(c) show, the chimera nature is also evident from the $\mbox{Re}(A(x,y,z))$ and $\mbox{Im}(A(x,y,z))$ fields. Using the local mean field $\tilde{\theta}(\mathbf{r})$, one can easily locate the unsynchronized filaments
\footnote{To identify the regions with unsynchronized phase, we consider the average of the absolute phase difference with its neighbors and select a suitable threshold.}.
An example is shown in Fig.~\ref{cap: cgle-knot}(a).

\subsection{CGLE: Minimum separation \& spontaneous fluctuations}

\begin{figure*}[p]

\subfigure[Snapshot of the filaments of a Hopf link. ]
{\begin{centering}
\includegraphics[width=0.4\columnwidth]{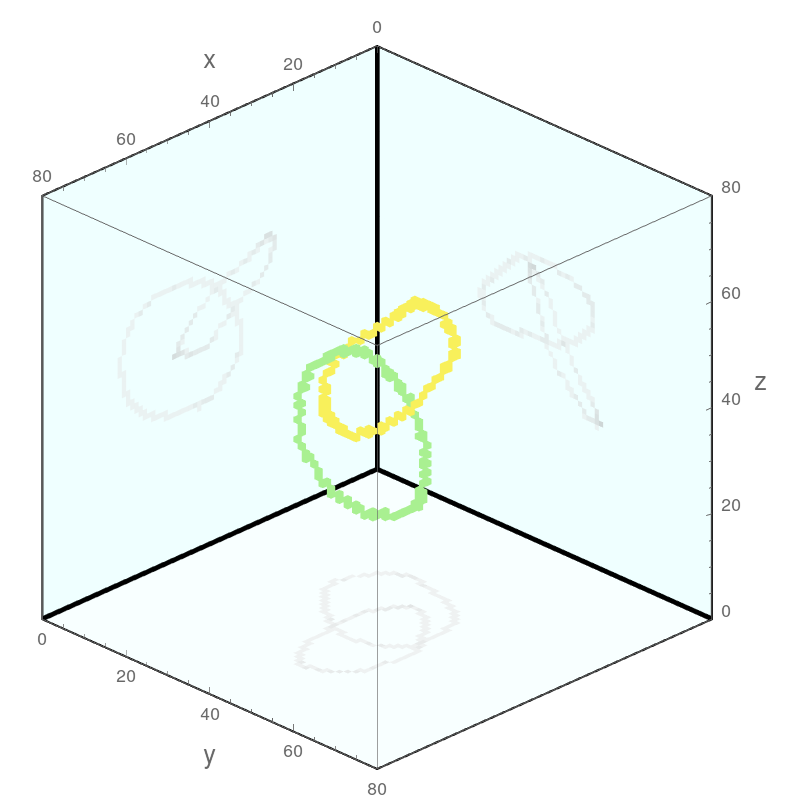}
\par\end{centering}
}
\qquad%
\subfigure[Snapshot of a straight filament with no-flux BC. ]
{\begin{centering}
\includegraphics[width=0.4\columnwidth]{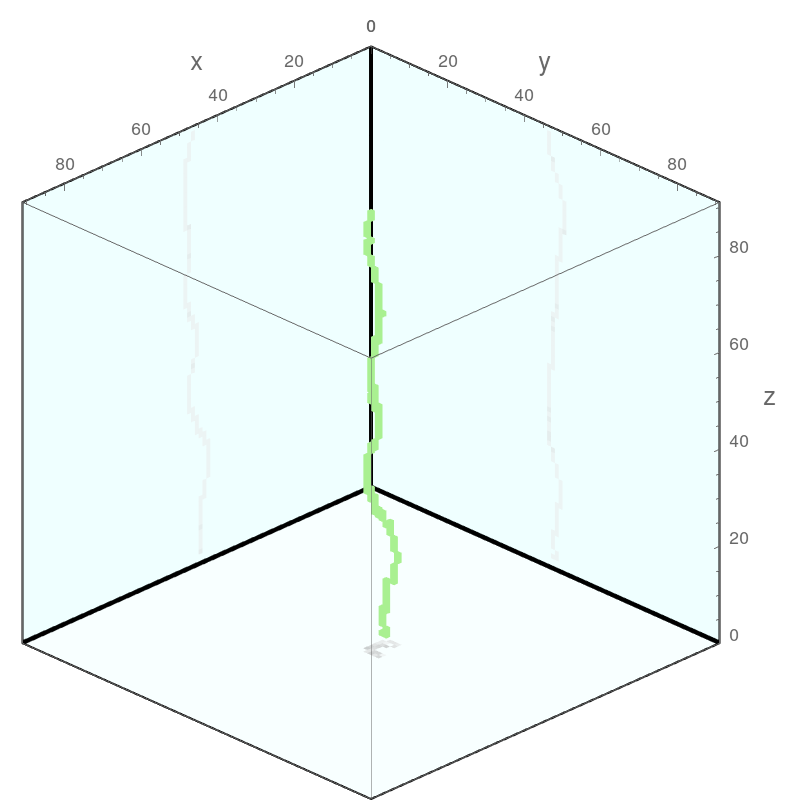}
\par\end{centering}
}

\subfigure[Circumferences $C_i$ of the two rings shown in (a) as a function of time.]
{\begin{centering}
\includegraphics[width=0.45\columnwidth]{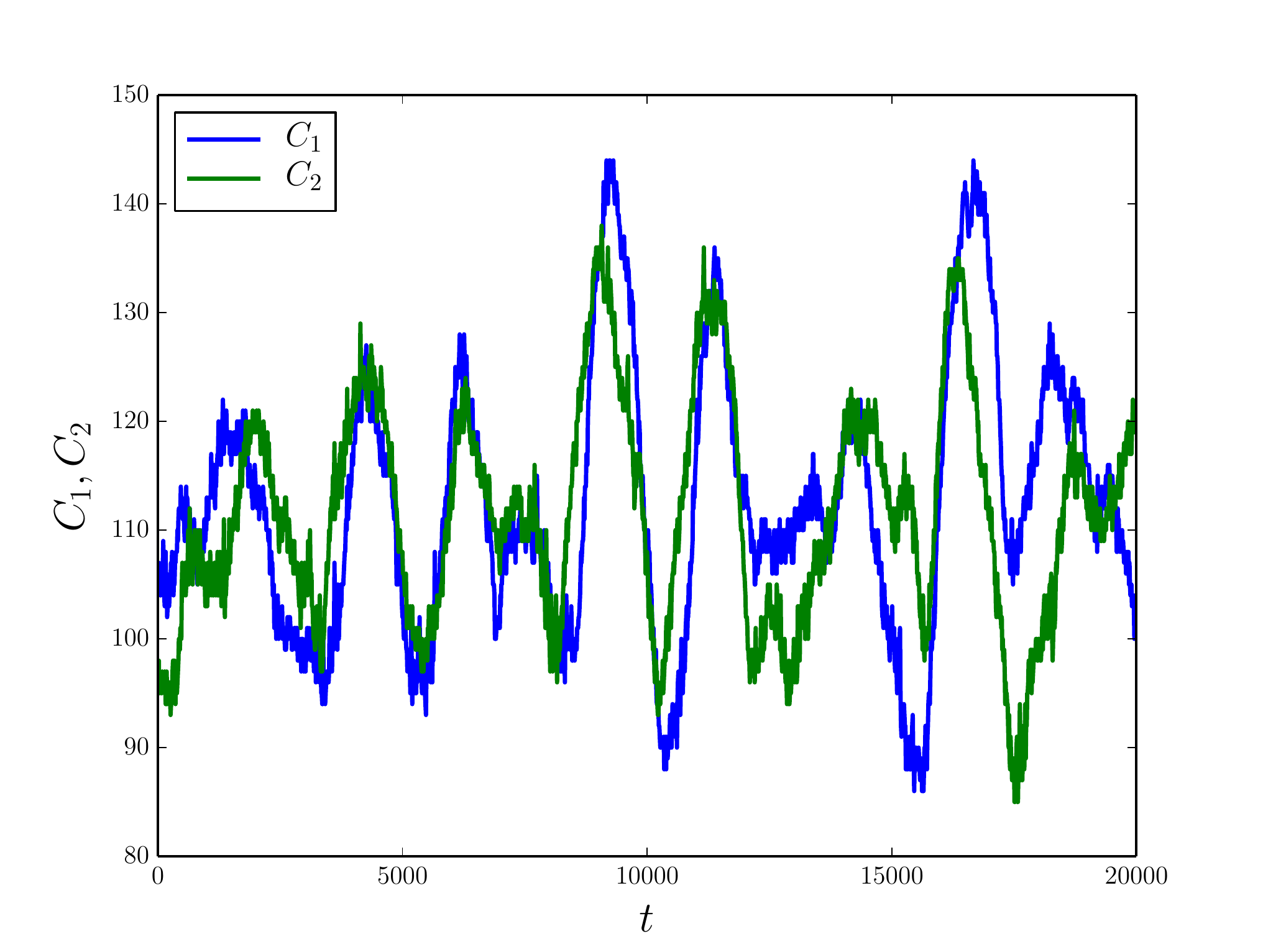}
\par\end{centering}
}
\qquad%
\subfigure[Length of the filament $\ell$ shown in (b) as a function of time. ]
{\begin{centering}
\includegraphics[width=0.45\columnwidth]{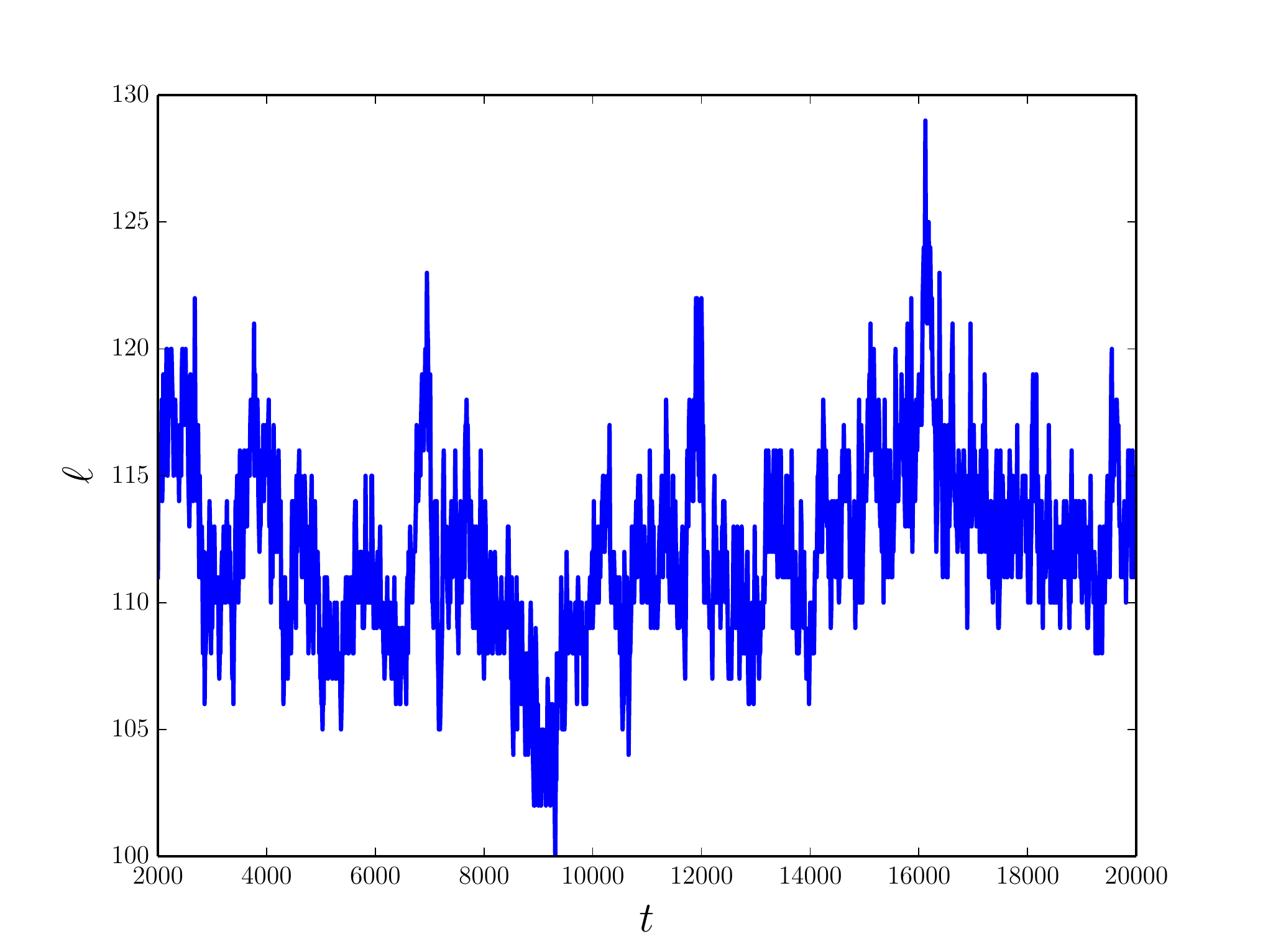}
\par\end{centering}
}

\begin{centering}
\subfigure[Minimum separation between the two rings shown in (a) as a function of time.]
{\begin{centering}
\includegraphics[width=0.45\columnwidth]{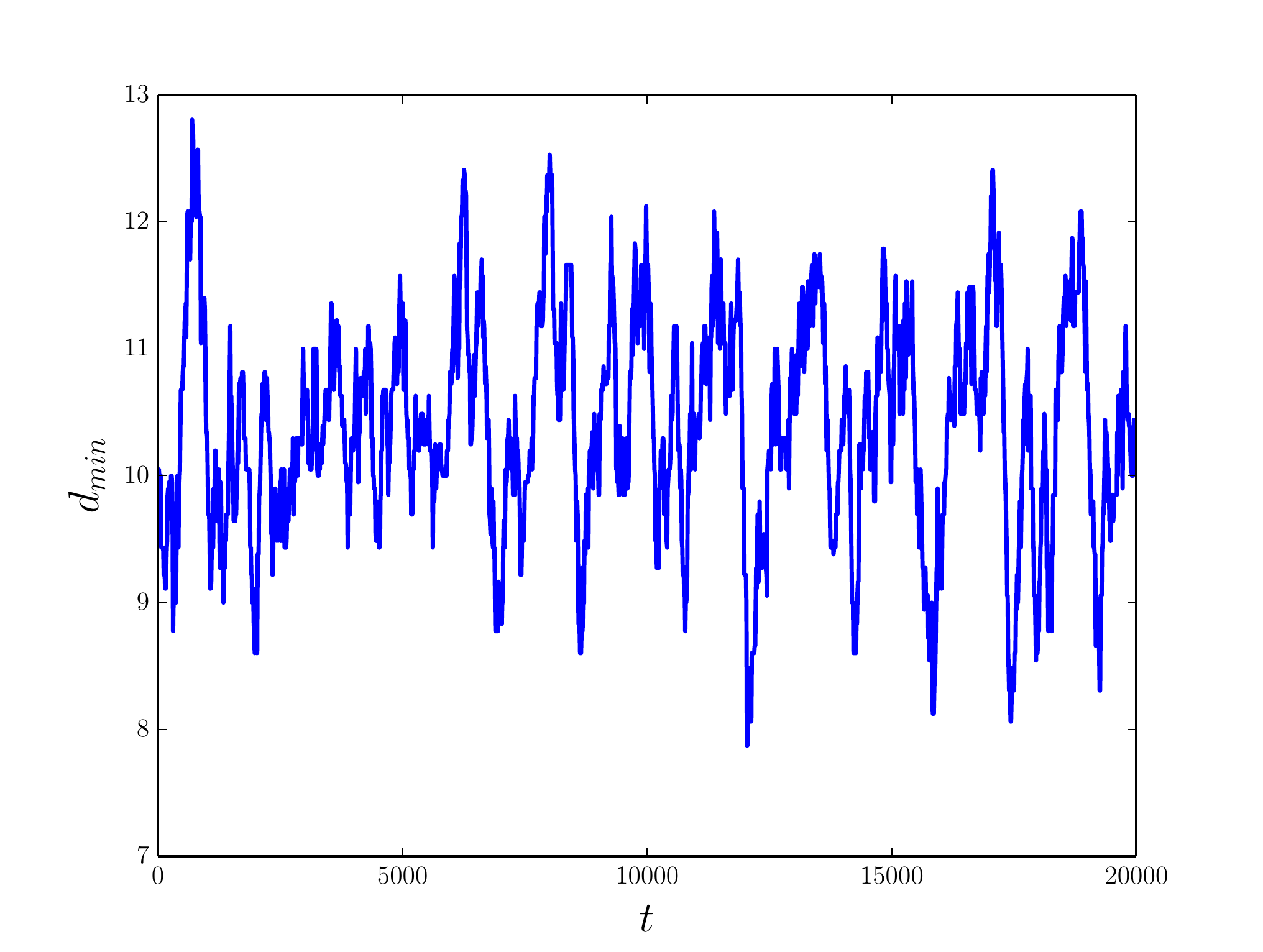}
\par\end{centering}
}
\qquad%
\subfigure[Roughness of the filament shown in (b) as a function of time.]
{\begin{centering}
\includegraphics[width=0.45\columnwidth]{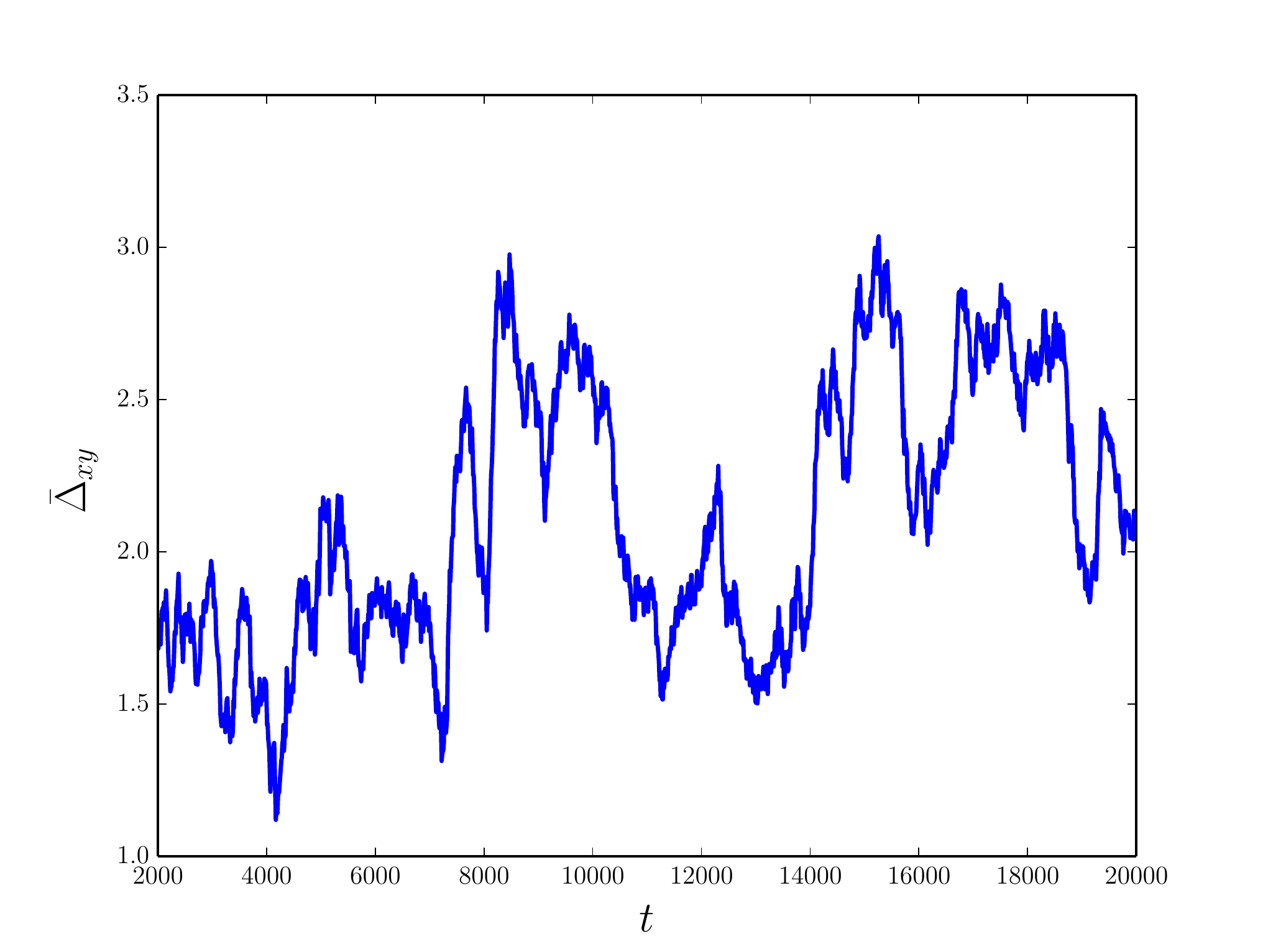}
\par\end{centering}
}

\par\end{centering}

\caption{\label{cap: method-knot-dmin}
CGLE with $R=4$, $a=1$, $b=0$, $K=0.1$. (left) Temporal evolution of a Hopf link with $L=80$. (right) Temporal evolution of a single filament oriented along the z-direction with $L=91$ which results in a time average filament length $\langle \ell \rangle \approx 110$ that is approximately the same as the time average circumference $\langle C \rangle \approx 110$ of the rings in the left column.
}
\end{figure*}

\begin{figure*}[p]

\subfigure[Average circumference and effective system size $L_{eff}$.]
{\begin{centering}
\includegraphics[width=0.44\columnwidth]{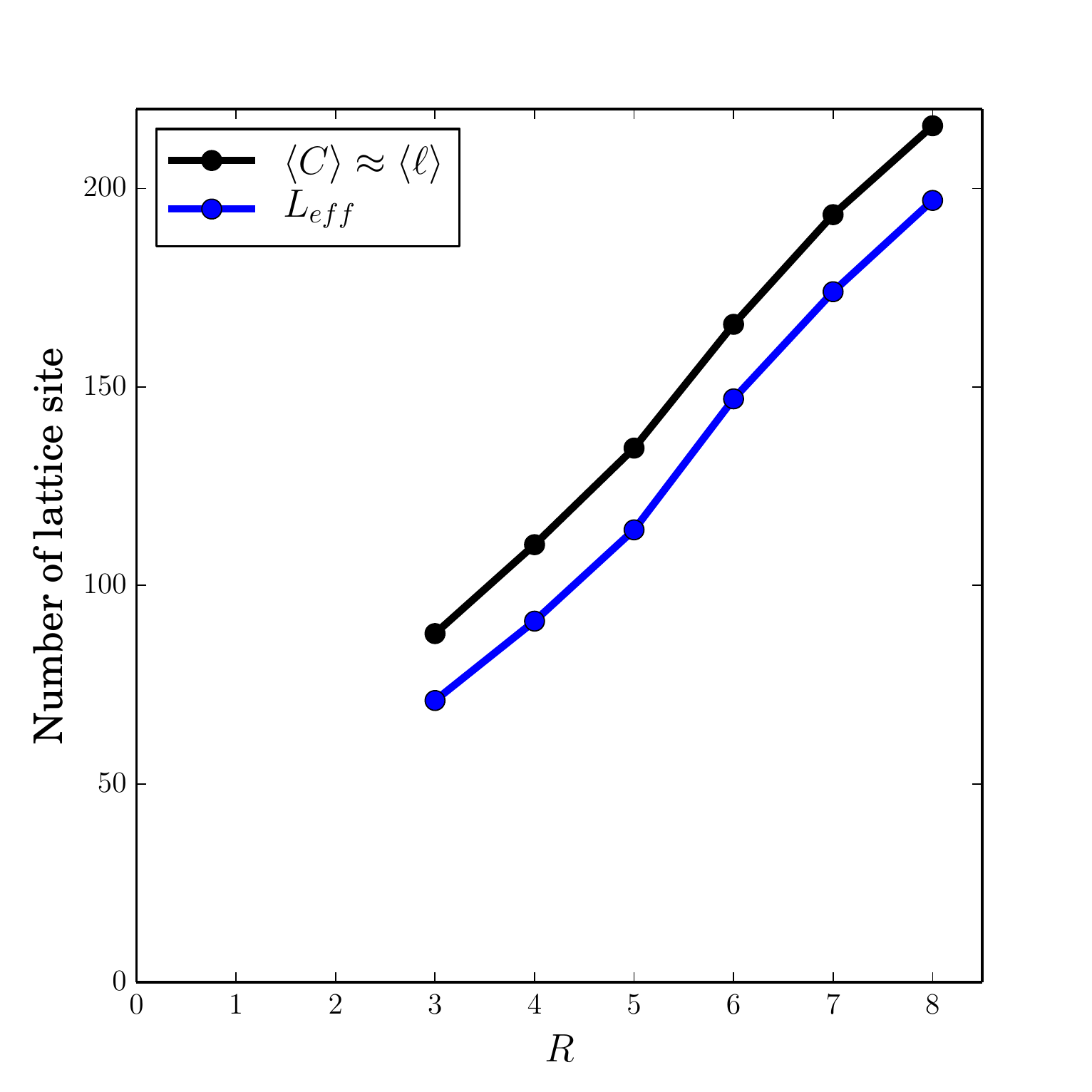}
\par\end{centering}
}
\subfigure[Comparing different length scale of fluctuation.]
{\begin{centering}
\includegraphics[width=0.54\columnwidth]{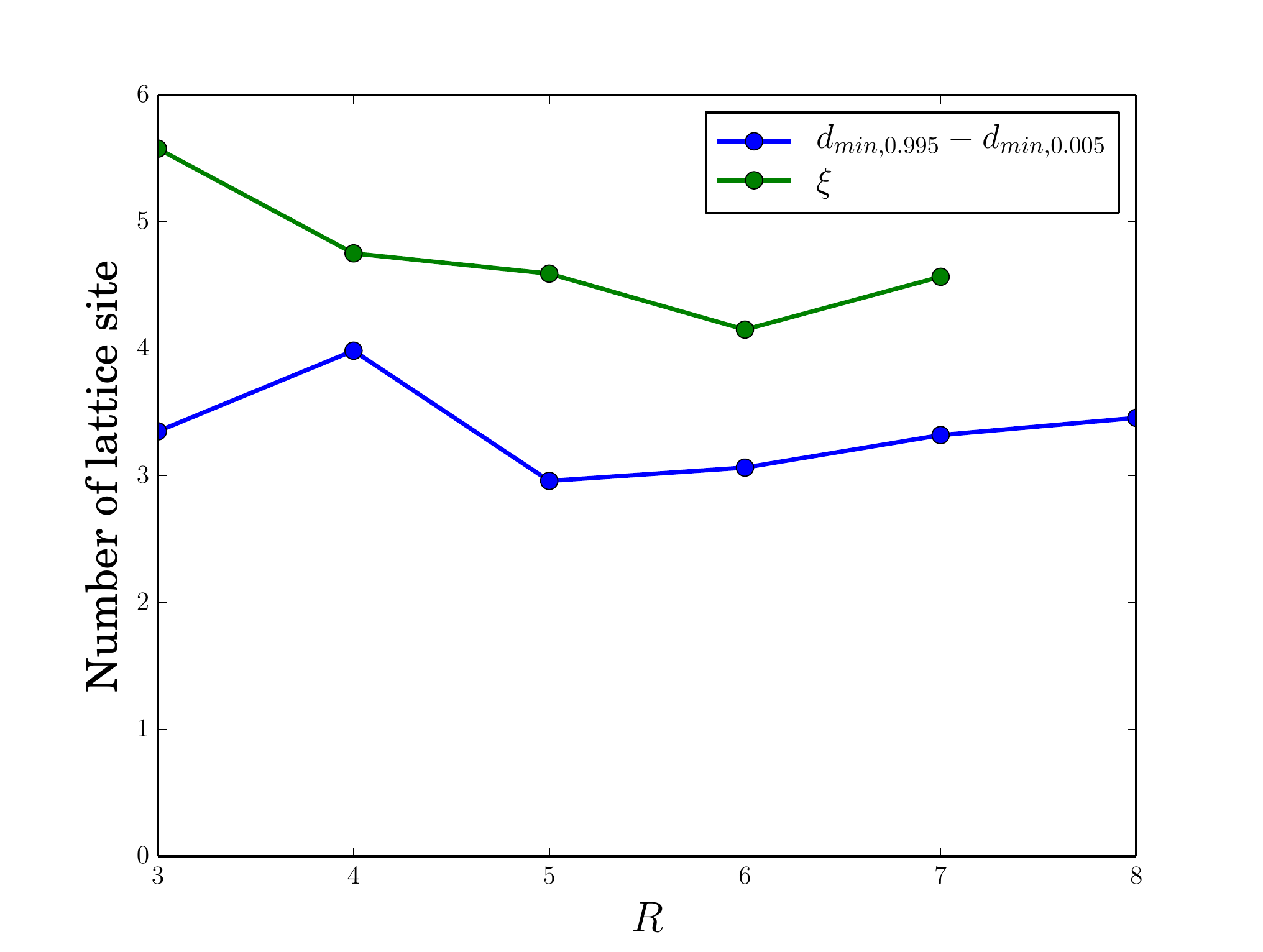}
\par\end{centering}
}

\caption{\label{cap: lenght-scale} 
CGLE with $a=1$, $b=0$, $K=0.1$ as in Fig.~\ref{cap: method-knot-dmin}.
(a) System size, $L_{eff}$, for which a single filament has the same average length $\langle \ell \rangle$ as the average circumference $\langle C \rangle$ of a Hopf link. (b) Measures of fluctuations for the case of a Hopf link (spread in the minimum separation between the rings, $d_{min,99.5\%}-d_{min,0.5\%}$, see Fig.~\ref{cap: method-knot-dmin}(e)) and for the case of a single filament (spread in the roughness, $\xi$, see Fig.~\ref{cap: method-knot-dmin}(f)), both as a function of $R$. 
}
\end{figure*}

When $R$ becomes too small, knots are no longer stable. This instability can be characterized by the dynamics of the filament(s) that make up the knots. Even though the region around the filament is unsychronized, the filaments can be found by a filament detection algorithm \cite{S_reid} of the mean field (see Fig. \ref{cap: method-knot-dmin}(a)). The length of filament can therefore be defined as the number of occupied lattice sites. Denote the two rings or filaments of a Hopf link as $F_1$ and $F_2$ with circumference (or length) $C_1$ and $C_2$, respectively. As Fig.~\ref{cap: method-knot-dmin}(c) shows, $C_1$ and $C_2$ fluctuate over time in a synchronous way. Fluctuations are also present in the minimum separation between $F_1$ and $F_2$, defined as $d_{min} = \min_{\textbf{r}_i\in F_i}(\textbf{r}_1,\textbf{r}_2)$, as shown in Fig. \ref{cap: method-knot-dmin}(e). To characterize these fluctuations statistically and identify an associated length scale, we consider the difference between the $99.5\%$-quantile and the $0.5\%$-quantile associated with $d_{min}$, corresponding to the error bars shown in Fig. 5(a) in the main paper. As shown in Fig. \ref{cap: lenght-scale}(b), this difference is not varying much across the considered values of $R$. This is in sharp contrast to the linear scaling of $d_{min}$ with $R$ (see Fig. 5(a) in the main paper). 

To substantiate that the intrinsic length scales associated with filament fluctuations do not strongly vary with $R$, we further consider the fluctuations of a single straight filament (see Fig.~\ref{cap: method-knot-dmin}(b)). To ensure a fair comparison with the fluctuations of Hopf links, we choose a {system size $L=L_{eff}$ such that the average single filament length $\langle\ell\rangle$ equals the average circumference $\langle C\rangle=(\langle C_1 \rangle + \langle C_2 \rangle)/2$ of the filaments in the Hopf link (see Fig.~\ref{cap: method-knot-dmin}(d)). The dependence of both these quantities as a function of $R$ is shown in Fig.~\ref{cap: lenght-scale}(a). To characterize the fluctuations of a single straight filament, we calculate its roughness. Due to the chosen initial conditions, the roughness is identical to the deviation from a straight filament oriented along the $z$-axis. Specifically, we define the deviation from the straight filament center $\bar{\textbf{r}}_{xy} = (1/L)\sum_z \textbf{r}_{xy}(z)$ to be
\begin{equation}
\Delta_{xy}(z) = |\textbf{r}_{xy}(z)-\bar{\textbf{r}}_{xy}|,
\end{equation}
where $\textbf{r}_{xy}(z)$ is the intersection point of the filament with the $x$-$y$ plane for a given $z$. The roughness $\bar{\Delta}_{xy} = (1/L)\sum_z \Delta_{xy}(z)$ is now simply $\Delta_{xy}(z)$ averaged over $z$.
As Fig.~\ref{cap: method-knot-dmin}(f) shows, the roughness varies over time. To characterize these (non-negative) fluctuations in the roughness over time and within an ensemble and to identify an associated length scale, we consider the $99\%$-quantile and denote it by $\xi$. This is the quantity shown in Fig. 5(a) in the main paper and again it does not vary much across the considered values of $R$. For a direct comparison with the length scale of fluctuations in the case of a Hopf link, please see Fig.~\ref{cap: lenght-scale}(b).

\subsection{Non-Local Rössler model}

The non-local Rössler model considered here is \cite{S_gu}:
\begin{eqnarray}
\dot{X}(\mathbf{r},t) & = & -Y-Z+K\int G(\mathbf{r}-\mathbf{r'})\left(X(\mathbf{r}')-X(\mathbf{r})\right)d\mathbf{r}',\\
\dot{Y}(\mathbf{r},t) & = & X+aY+K\int G(\mathbf{r}-\mathbf{r'})\left(Y(\mathbf{r}')-Y(\mathbf{r})\right)d\mathbf{r}',\\
\dot{Z}(\mathbf{r},t) & = & b+Z(X-c),
\end{eqnarray}
where the control parameters are $(a,b,c)$ and the coupling strength is
$K$. Again, we can use $\ensuremath{(X,Y,Z)=(\cos\theta,\sin\theta,0)}$
with $\theta(\mathbf{r})$ from states with knotted structures generated by the Kuramoto model as IC. 
When $a=b=0.2$, the effective $|\alpha|$ decreases as $c$ increases \cite{S_gu-thesis}. We observe stable knots within $3.3\lesssim c\lesssim5$ for weak coupling $K=0.05$ provided that $L \gg R \gg 1$. Note that as $c$ increases the intrinsic dynamics of the oscillators also changes. Namely, the dynamics undergoes a period-doubling cascade to chaotic oscillations. In particular, we observe stable knots in the period-2 regime with $c=3.6$ (Fig.~\ref{cap: rossler-knot}), as well as in the chaotic regime with $c=4.8$ (Fig.~\ref{cap: rossler-knot-chaotic}). 

An additional feature of the wave dynamics in these regimes is evident from Fig.~\ref{cap: rossler-knot} and Fig.~\ref{cap: rossler-knot-chaotic}: The amplitudes are modulated. For example, in the period-2 regime alternating wave maxima are present. A topological consequence of such a behavior is that two dimensional structures exist such that the local dynamics has a lower period than that of the bulk. Specifically, these structures, called synchronization defect sheets (SDSs) in the following, separate domains of different oscillation phases and for periodic BC either originate from a filament or are closed. More importantly, every filament has an attached SDS such that they become part of any knotted structure. This can already be observed in the cross-sections shown in Fig.~\ref{cap: rossler-knot} and Fig.~\ref{cap: rossler-knot-chaotic}. To clearly identify SDSs, we use the detection algorithm developed for the lower dimensional case in Ref.~\cite{S_gu-thesis}. A specific example of SDSs is shown in Fig.~\ref{cap: rossler-SDS} and their subsequent motion is shown in the Supplementary Video.

\begin{figure*}[p]
\begin{centering}
\subfigure[Hopf link]{\begin{centering}
\includegraphics[width=0.45\columnwidth]{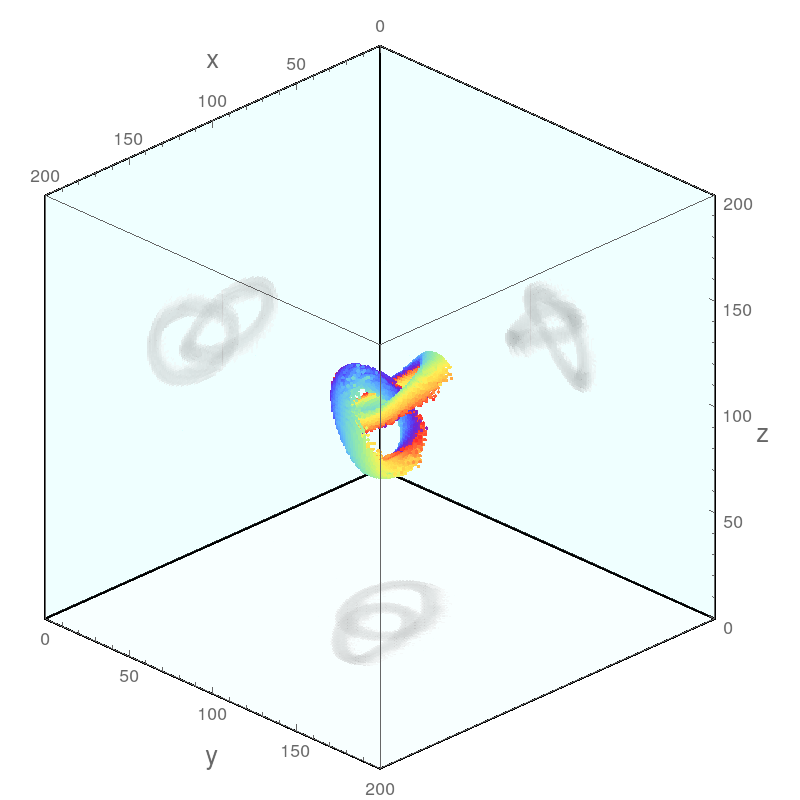}
\par\end{centering}

}
\par\end{centering}

\begin{centering}
\subfigure[$X$]{\begin{centering}
\includegraphics[width=0.33\columnwidth]{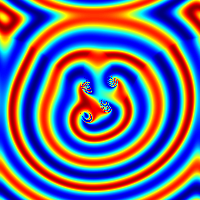}
\par\end{centering}

}\subfigure[$Y$]{\begin{centering}
\includegraphics[width=0.33\columnwidth]{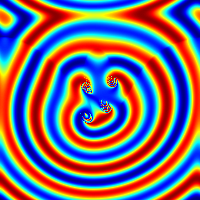}
\par\end{centering}

}
\par\end{centering}

\begin{centering}
\subfigure[$Z$]{\begin{centering}
\includegraphics[width=0.33\columnwidth]{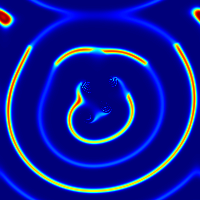}
\par\end{centering}

}\subfigure[$\theta=\tan^{-1}(Y/X)$]{\begin{centering}
\includegraphics[width=0.33\columnwidth]{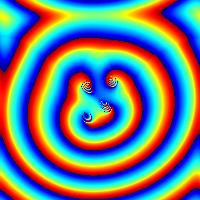}
\par\end{centering}

} 
\par\end{centering}

\caption{\label{cap: rossler-knot} Snapshot of a Hopf link in the non-local R{\"o}ssler model for $(a,b,c,K)=(0.2,0.2,3.6,0.05)$, corresponding to the period-2 regime. The lifetime of the knot is $\tau>10^{5}$. The same 2D cross-sections of the Hopf link are shown in (b-e) for the different fields $X$, $Y$, $Z$ and $\theta$. The color scheme is such that deep blue represents the most negative value, and red represents the most positive value. The discontinuities of color along the wave fronts correspond to cross-sections of synchronization defect sheets. $L=200$, $R=8$ and periodic BC.}
\end{figure*}
\begin{figure*}[p]
\begin{centering}
\subfigure[Hopf link]{\begin{centering}
\includegraphics[width=0.45\columnwidth]{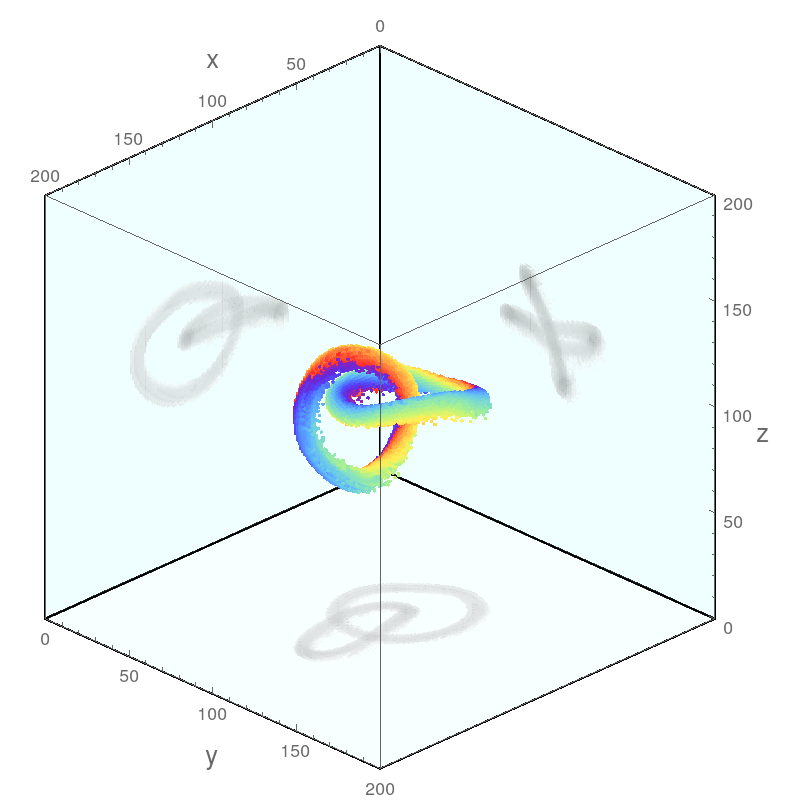}
\par\end{centering}

}
\par\end{centering}

\begin{centering}
\subfigure[$X$]{\begin{centering}
\includegraphics[width=0.33\columnwidth]{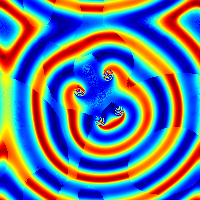}
\par\end{centering}

}\subfigure[$Y$]{\begin{centering}
\includegraphics[width=0.33\columnwidth]{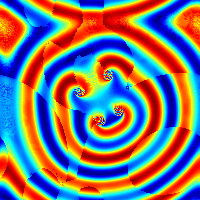}
\par\end{centering}

}
\par\end{centering}

\begin{centering}
\subfigure[$Z$]{\begin{centering}
\includegraphics[width=0.33\columnwidth]{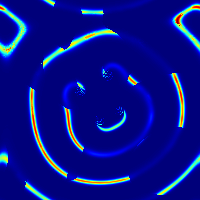}
\par\end{centering}

}\subfigure[$\theta=\tan^{-1}(Y/X)$]{\begin{centering}
\includegraphics[width=0.33\columnwidth]{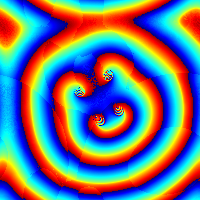}
\par\end{centering}

} 
\par\end{centering}

\caption{\label{cap: rossler-knot-chaotic} Similar to Fig. \ref{cap: rossler-knot},
but in the chaotic regime $(a,b,c,K)=(0.2,0.2,4.8,0.05)$. }
\end{figure*}
\begin{figure*}[p]
\begin{centering}
\subfigure[Synchronization defect sheets]{\begin{centering}
\includegraphics[width=0.5\columnwidth]{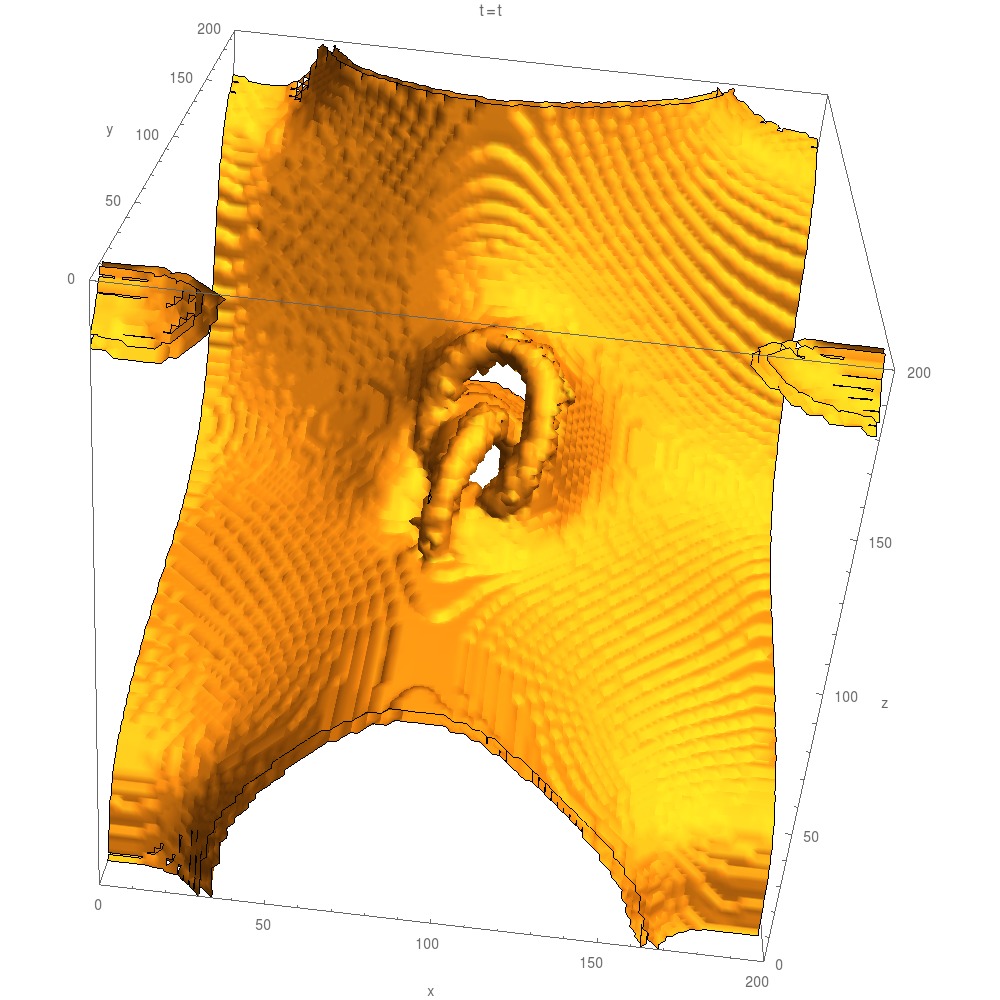}
\par\end{centering}

}\subfigure[Synchronization defect sheets (another view)]{\begin{centering}
\includegraphics[width=0.5\columnwidth]{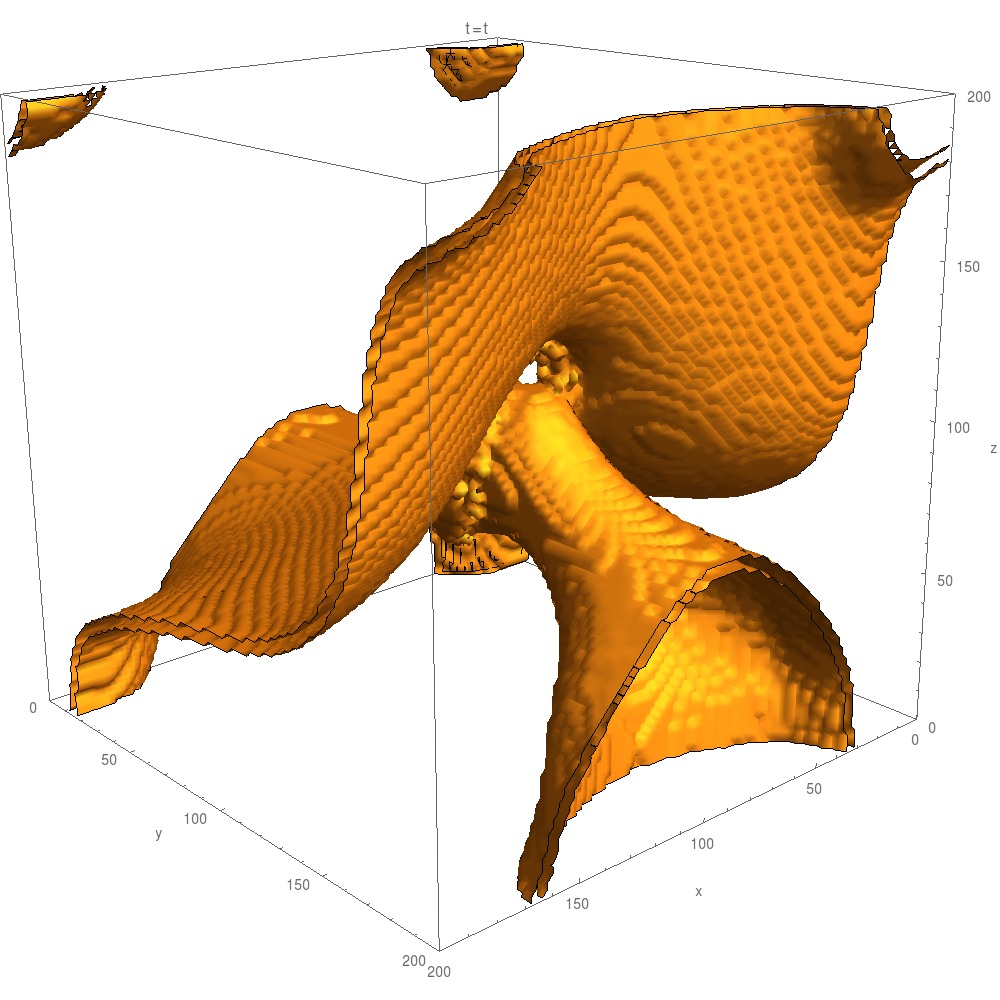}
\par\end{centering}

}
\par\end{centering}

\begin{centering}
\subfigure[$z=50$]{\begin{centering}
\includegraphics[width=0.3\columnwidth]{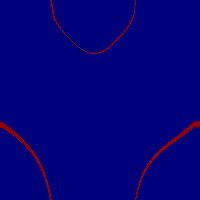}
\par\end{centering}

}\subfigure[$z=85$]{\begin{centering}
\includegraphics[width=0.3\columnwidth]{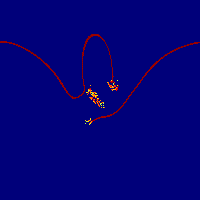} 
\par\end{centering}

}\subfigure[$z=93$]{\begin{centering}
\includegraphics[width=0.3\columnwidth]{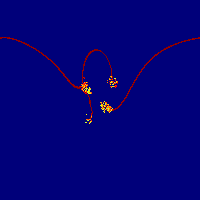} 
\par\end{centering}

}
\par\end{centering}

\begin{centering}
\subfigure[$z=109$]{\begin{centering}
\includegraphics[width=0.3\columnwidth]{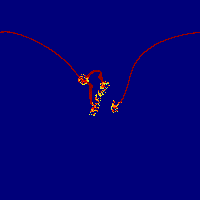} 
\par\end{centering}

}\subfigure[$z=118$]{\begin{centering}
\includegraphics[width=0.3\columnwidth]{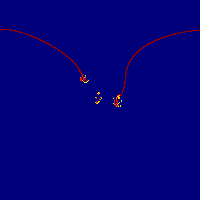} 
\par\end{centering}

}\subfigure[$z=130$]{\begin{centering}
\includegraphics[width=0.3\columnwidth]{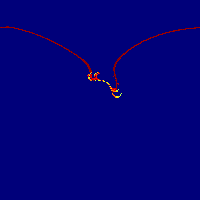} 
\par\end{centering}

}
\par\end{centering}

\caption{\label{cap: rossler-SDS} Visualization of the synchronization defect sheets (SDSs) present in Fig.~\ref{cap: rossler-knot}. (a), (b): Different 3D plots of the SDSs. (c)-(h): 2D cross-sections at different values of $z$. The red lines represent the cross-sections of SDSs and the yellow-red dots indicate the unsynchronized regions. Note that the cross-section in (e) is the same cross-section as in Fig.~\ref{cap: rossler-knot} (b-e).}
\end{figure*}

\end{document}